\documentclass[14pt,a4paper]{article}
\usepackage{amsmath}
\usepackage[dvips]{graphicx}
\usepackage{setspace}

\usepackage[numbers,compress]{natbib}%

\usepackage[margin=2.5cm]{geometry}
\usepackage{epsfig}
\usepackage{subfigure}
\usepackage[section]{placeins}
\title{\bf SOME ASPECTS OF OPTICAL SPATIAL SOLITONS  IN PHOTOREFRACTIVE CRYSTALS}

\date{11 September, 2012}
\author{ S. Konar$^{1}$  and Anjan Biswas$^{2}$\\
$^{1}$Department of Applied Physics, Birla Institute of Technology,
Mesra, Ranchi-835215,\\ Jharkhand, India\\ \\$^{2}$Department of
Mathematical Sciences, Delaware State University,\\ Dover, DE
19901-2277, USA }

\begin{document}
\maketitle

\setcounter{page}{1}

\begin{onehalfspacing}
\begin{abstract}

 { We have reviewed recent developments of some aspects
 of optical spatial solitons in photorefractive media. Underlying principles
 governing the dynamics of photorefractive nonlinearity have been
 discussed using band transport model. Nonlinear dynamical equations
 for propagating solitons
 have been derived considering single
as well as two-photon photorefractive processes. Fundamental
properties of three types of solitons, particularly, screening,
photovoltaic and screening photovoltaic solitons have been
considered. For each type of solitons, three different
configurations i.e., bright, dark and gray varieties have been
considered. Mechanisms of formation of these solitons due to single
as well as two-photon photorefractive processes have been considered
and their self bending discussed. Vector solitons, particularly,
incoherently coupled solitons due to single photon and two-photon
photorefractive phenomena have been highlighted. Existence of some
missing solitons have been also pointed out. }

\end{abstract}\ \ \ \ \ \ \ \ \ \ \ \ \ \ \ \ \ \ \ \ \ \
\end{onehalfspacing}
\setcounter{page}{1}
\ \ \ \ \ \ \ \ \ \ \ \ \ \ \ \ \ \ \ \ \ \ \ \ \ \ \
\newpage
\tableofcontents

\newpage
\section{Introduction}

\begin{onehalfspacing}
\setcounter{equation}{0} \ \ \ \
 The advent of nonlinear optics has paved the way to a number of fundamental
 discoveries. Exotic one, among many of these, is the discovery of optical solitons ~\citep{1,2,3,4,5,6,7,8,9,10,11,12,13,14,15,16}.
 These are optical pulses or beams which are able to propagate without broadening and
 distortion. Optical solitons are envelope solitons, and
the term soliton was first coined by Zabusky and Kruskal in 1965 to
reflect the particle like nature of these waves that remain intact
even after mutual collision ~\citep{4}.
 Optical solitons are extensively studied topic not solely due to their mathematical
 and physical elegance but as well as due to the possibility of real life applications.
   They have been contemplated as the building blocks of soliton based optical
   communication systems, signal processing, all optical switching, all optical devices
   etc.\\

\ \ \ \ \ In optics, three different types of solitons are known
till date, these are, temporal ~\citep{2,3},
   spatial~\citep{5,6}  and spatio-temporal
   solitons~\citep{7,8}. Temporal solitons are short optical pulses which maintain their temporal shape
   while propagating over long distance. The way a temporal optical soliton is established
is that a nonlinear pulse sets out in dispersive medium and develops
a chirp. Then the dispersion produces a chirp of opposite sign. A
temporal soliton pulse results due to the balancing of these
opposite chirps across the width of the pulse, which arise from the
material dispersion and nonlinearity.These opposite chirps balance
each other  when dispersion is completely canceled by the
nonlinearity of the medium. Temporal solitons are routinely
generated in optical fibers~\citep{2}
   and they are backbone of soliton based optical communication systems, soliton lasers etc. In contrast, optical spatial solitons are
   beams of electromagnetic energy that rely upon balancing diffraction and nonlinearity to retain their
shape.  While propagating in the nonlinear medium,
   the optical beam modifies the refractive index of the medium in such a way that the spreading due
   to diffraction is eliminated. Thus, optical beam induces a nonlinear waveguide and at the same
   time guided in the waveguide it has induced. This means soliton is a guided mode of the nonlinear waveguide induced by
   it.\\

Though temporal solitons can be easily generated in optical fibers,
generation of spatial optical solitons is a much more difficult
task. For example, in silica glass, the nonlinearity is proportional
to light intensity and  the value of the proportionality constant
$n_{2}$   is of the order of  $10^{-16}cm^{2}/W$     only.
Therefore, in order to compensate for the beam spreading due to
diffraction, which is a large effect, required optical nonlinearity
is very large, and, consequently optical power density is also
large~\citep{8}. Another impediment in detecting spatial solitons,
in bulk Kerr nonlinear media, is the catastrophic collapse of the
optical beam, which is inevitable with  Kerr nonlinearity. Discovery
of non Kerr nonlinearities, whose  mechanism is different from Kerr
nonlinearity, has lead to the revelation of stable three dimensional
soliton formations without catastrophic collapse. These
nonlinearities are photorefractive nonlinearity~\citep{10,18},
quadratic nonlinearity~\citep{19,20,21,22}  and resonant electronic
nonlinearity in atoms or molecules~\citep{23,24,25,26,27}. With the
identification of photorefractive nonlinearity, which possesses
strong nonlinear optical response, it is possible to create optical
solitons at very low light intensity. Spatial photorefractive
optical solitons possess some unique properties which make them
attractive in several applications, such as, all optical switching
and routing, interconnects, parallel computing, optical storage
etc~\citep{15,16,28,35}. They are also promising for experimental
verification of theoretical models, since, they can be created at
very low power. In the present article, we confine our discussion on
the properties of photorefractive spatial solitons.

\section{Photorefractive Effect}

The photorefractive(PR) effect is the change in refractive index of
certain electro-optic materials owing to optically induced
redistribution of charge carriers. Light induced refractive index
change occurs owing to the creation of space charge field, which is
formed due to nonuniform light intensity. Originally the
photorefractive effect was considered to be undesirable, since this
leads to scattering and distortion of collimated optical
beams~\citep{36}. Soon it was realized that these materials have
potential applications in holography~\citep{33}, optical phase
conjugation~\citep{32}, optical signal processing and optical
storage~\citep{34,35}. Photorefractive materials are classified in
three different categories. Most commonly used photorefractive
materials are inorganics,  such as, $LiNbO_{3}$, $BaTiO_{3}$,
$Sr_{x}Br_{1-x}Nb_{2}O_{6}$,  $KNbO_{3}$, $Bi_{12}SiO_{20}$,
$Bi_{12}TiO_{20}(BTO)$ etc. Semiconductors have large carrier
mobility that produces fast dielectric response time, which is
important for fast image processing. Therefore, photorefractive
semiconductors, such as, GaAs, InP, CdTe etc., complement the
photorefractive ferroelectrics with the potential of fast
holographic processing of optical information. Polymers also show
strong photorefractive effect~\citep{37,38,39,40}. They are easy to
produce and PR effect in polymers appear only if a high voltage is
applied. Strong photorefractive pattern can be erased easily in
polymers by decreasing the applied voltage. Polymers  show good
temperature stability, and for a given applied voltage, they usually
show stronger refractive index change in comparison to inorganic
crystals with equal doping densities.

\section{Origin of Photorefractive Nonlinearity}
 In a photorefractive material, the spatial distribution of
intensity of the optical field gives rise to an inhomogeneous
excitation of charge carriers. These charge carriers migrate due to
drift and or diffusion and produce a space charge field, whose
associated electric field modifies the refractive index of the
crystal via Pockel's effect~\citep{32,34,35}. For a
noncentrosymmetric photorefractive crystal, the refractive index
change due to the linear electro-optic effect is given
by~\citep{32,34}

    \begin{equation}\label{e1}
    \Delta n =-\frac{1}{2}n^{3}r_{eff}E_{sc} ,
   \end{equation}
    where $n$ is the average refractive index, $r_{eff}$ is the effective
    linear electro-optic coefficient which depends on the orientation of the
    crystal and polarization of light, and $E_{sc}$ is the space charge field.
A unique feature of photorefractive materials is their ability to
exhibit both self focusing and  defocusing nonlinearity in the same
crystal. This is achieved by changing the polarity of the biased
field, which in turn changes the polarity of the space charge field
$E_{sc}$. Hence, the same crystal can be used to generate either
bright ( require self focusing nonlinearity) or dark and gray
solitons ( require defocusing nonlinearity). Photorefractive
nonlinearity is also wavelength sensitive, thus, it is possible to
generate solitons at one wavelength and then use the soliton
supported channel to guide another beam at different wavelength.

\section{Band Transport Model}

 The most widely accepted theoretical
formulation of photorefractive phenomenon is described by
Kukhtarev-Vinetskii band transport model~\citep{41}. A schematic
diagram of this model is shown in figure (1), where a PR crystal is
being illuminated by an optical beam with nonuniform intensity.
\begin{center}
\textit{Insert Figure (1) here}
\end{center}

The photorefractive medium, at the ground level, has completely full
valance band and an empty conduction band. The material has both
donor and acceptor centers, uniformly distributed, whose energy
states lies somewhere in the middle of the band gap. The acceptor
electron states are at much lower energy in comparison to that of
the donor electron states. The presence of a nonuniform light beam
excites unionized donor impurities, creates charge carriers which
move to the conduction band where they are free to move  to diffuse
or to drift under the combined influence of self generated and
external electric field and are finally trapped by the acceptors.
During this process, some of the electrons are captured by ionized
donors and thus are neutralized. In the steady state, the process
leads to the charge separation, which tends to be positive in the
illuminated
region and vanish in the dark region.\\\

\ \ \ \ \ We assume that the donor impurity density is $N_{D}$ and
acceptor impurity density $N_{A}$. If the ionized donor density is
$N_{D}^{+}$, then the rate of electron generation due to light and
thermal processess is $(s_{i}I+\beta_{T})(N_{D}-N_{D}^{+}),$ where
$s_{i}$ is the  cross section of photoexcitation, $I$ is the
intensity of  light which is written in terms of Poynting flux
$I=\frac{n}{2\eta_{0}}|\Phi|^{2}$  ,  $\Phi$  is the electric field
of light, $\epsilon _{0}$ is the free space permittivity, and
$\beta_{T}$ is the rate of thermal generation. If $N $ is the
electron density and $\gamma _{R}$ the electron trap recombination
coefficient, then, the rate of recombination of ionized donars with
free electrons is $\gamma _{R}N N_{D}^{+}$. Thus, the rate equation
describing the donar ionization  is given by

\begin{equation}\label{e2}
\frac{\partial N_{D}^{+}}{\partial
t}=(s_{i}I+\beta_{T})(N_{D}-N_{D}^{+})-\gamma _{R}N N_{D}^{+}.
\end{equation}

The electron concentration is affected by  recombination with
ionized donors and due to migration of electrons, resulting in an
electron current with current density $J$, hence, electron
continuity equation turns out to be

\begin{equation}\label{e3}
\frac{\partial N}{\partial t}-\frac{\partial N_{D}^{+}}{\partial t}
=\frac{1}{e}\overrightarrow{\nabla }.\overrightarrow{J},
\end{equation}

and

\begin{equation}\label{e4}
\overrightarrow{J}=e N\mu \overrightarrow{E} +k_{B}T\mu
\overrightarrow{\nabla }N +
k_{p}s_{i}\left(N_{D}-N_{D}^{+}\right)I\vec{c},
\end{equation}

where the current density $\overrightarrow{J}$ is the sum of the
contributions from the drift, diffusion and photovoltaic effect; $e$
is the electronic charge, $\mu$  is the electron mobility, $k_{B}$
is the Boltzmann constant, $T$ is electron temperature,  $k_{p}$ is
the photovoltaic constant and $\vec{c}$ is the unit vector in the
direction of c-axis of the PR crystal. $E$ is the total electric
field including the one externally applied and that associated with
the generated space charge. The redistribution of electrical charges
and the creation of space charge field obey the Poisson's equation,
therefore,
\begin{equation}\label{e5}
\overrightarrow{\nabla}\cdot\epsilon \overrightarrow{E}=\rho ,
\end{equation}

and the charge density   $\rho$   is given by
\begin{equation}\label{e6}
\rho
\left(\overrightarrow{r}\right)=e\left(N_{D}^{+}-N_{A}-N\right).
\end{equation}\
Equations (2) -(6) can be solved to find out the space charge field
$E_{sc}$ and subsequently the optical nonlinearity in the
photorefractive media.\

\section{Space Charge Field}

To estimate the nonlinear index change in photorefractive media due
to the presence of nonuniform optical field, we need to calculate
the screening electric field $E_{sc}$. The response of a
photorefractive material to the applied  optical field is
anisotropic and it is nonlocal function of  light intensity.
Anisotropy does not allow radially symmetric photorefractive
solitons~\citep{42,43}. To formulate a simple problem, and since,
most of the experimental investigations on photorefractive solitons
are one dimensional waves, it is appropriate to find material
response in one dimension ( say x only). Steady state
photorefractive solitons may be obtained under homogeneous
background illumination, which enhances  dark conductivity of the
crystal. In the steady state, induced space charge field $E_{sc}$
can be obtained from the set of rate, continuity, current equations
and Gauss law. In the steady state, and in one dimension, these
equations reduce to~\citep{44,45,46}:

\begin{equation}\label{e7}
s_{i}(I+I_{d})(N_{D}-N_{D}^{+})-\gamma_{R}NN_{D}^+=0,
\end{equation}

\begin{equation}\label{e8}
\frac{\partial E_{sc}}{\partial
x}=\frac{e}{\epsilon_{0}\epsilon_{r}}(N_{D}^{+}-N_{A}-N),
\end{equation}

\begin{equation}\label{e9}
J=e N\mu E_{sc} +k_{B}T\mu \frac{\partial N}{\partial x} +
k_{p}s_{i}\left(N_{D}-N_{D}^{+}\right)I,
\end{equation}

\begin{equation}\label{e10}
\frac{\partial J}{\partial x}=0,
\end{equation}

where  $\epsilon_{r}$  is the relative static permittivity; $I_{d}
(=\frac{\beta_{T}}{s_{i}})$ is the so called dark irradiance that
phenomenologically accounts for the rate of thermally generated
electrons. This is also the homogeneous intensity that controls the
conductivity of the crystal. In most cases, the optical intensity
$I$ is such that for electron dominated photo-refraction,    $N\ll
N_{D}$,  $N\ll N_{A}$ and $N_{A}\ll N_{D}^{+}$. Under this usually
valid situation, the space charge field $E_{sc}$  is related to the
optical intensity $I$ through

\begin{equation}\label{e11}
E_{sc}=E_{0}\frac{I_{\infty} +I_{D}}{I+I_{D}}+E_{p}\frac{I_{\infty}
-I}{I+I_{D}}-\frac{k_{B}T}{e}\frac{1}{I+I_{D}}\frac{\partial
I}{\partial x},
\end{equation}

where  $E_{o}$ is the external bias field to the photorefractive
crystal,   $E_{p}=\frac{k_{p} \gamma_{R} N_{A}}{e\mu}$ is the
photovoltaic field. In addition, we have assumed that the power
density of the optical field attains asymptotically a constant value
at $x \rightarrow \pm\infty$  i.e., $I(x \rightarrow
\pm\infty,z)=I_{\infty}$. Moreover, in the region of constant
illumination, equations(7)-(10) require that the space charge field
$E_{sc}$ is independent of  $x$ i.e., $E_{sc}(x \rightarrow
\pm\infty,z)=E_{o}$.

\section{Photorefractive Nonlinearity }

The space charge induced change  $\triangle n$   in the refractive
index $n$ is obtained as

\begin{equation}\label{e12}
|\triangle n|=\frac{1}{2}n^{3}r_{eff}\left[E_{0}\frac{I_{\infty}
+I_{D}}{I+I_{D}}+E_{p}\frac{I_{\infty}
-I}{I+I_{D}}-\frac{k_{B}T}{e}\frac{1}{I+I_{D}}\frac{\partial
I}{\partial x}\right].
\end{equation}\

It is evident from above expression that the change in refractive
index is intensity dependent, i.e., under the action of nonuniform
illumination the photorefractive crystal has become an optical
nonlinear medium. The first term in the above equation is generally
known  as the screening term. In nonphotovoltaic photorefractive
material, this is the main term which is responsible for soliton
formation. The space charge redistribution in a photorefractive
crystal is caused mainly by the drift of photoexcited charges under
a biasing electric field. This mechanism leads directly to a local
change of refractive index and is responsible for self-focusing of
optical beams. The second term is due to photovoltaic effect,  which
leads to the formation of photovoltaic solitons. First two terms do
not involve spatial integration, therefore, they are local nonlinear
terms. Functional form of both terms are different from Kerr
nonlinearity.  These terms are similar to that of saturable
nonlinearity and explain why the collapse phenomenon is not
observable  in photorefractive materials. Besides drift mechanism,
transport of charge carriers occurs also due to diffusion process.
This process results in the nonlocal contribution to the refractive
index change. The last term is due to diffusion. The strength of the
diffusion effect is determined by the width of the soliton forming
optical beam. In case of strong bias field and relatively wide
beams, the diffusion term is often neglected. However, its
contribution can become significant for very narrow spatial solitons
or optical beams. When diffusion is significant, it is responsible
for deflection of photorefractive solitons~\citep{44,45,46}.

\section{Classification of Photorefractive Solitons}

\subsection{Solitons Due to Single Photon Photorefractive Phenomenon }

Till date, three different types of steady state spatial solitons
have been predicted in photorefractive media, which owe their
existence due to single photon photorefractive phenomenon.
Photorefractive screening solitons were identified first. In the
steady state, both bright and dark screening solitons (SS) are
possible when an external bias voltage is appropriately applied to a
non-photovoltaic photorefractive
crystal~\citep{47,48,49,50,51,52,53,54,55,56}. Intuitively, the
formation of bright screening photorefractive solitons can be
understood as follows. When a narrow optical beam propagates through
a biased photorefractive crystal, as a result of illumination,
conductivity in the illuminated  region increases and the
resistivity decreases. Since the resistivity is not uniform across
the crystal, the voltage drops primarily across the non illuminated
region and voltage drop is minimum in the high intensity region.
This leads to the formation of large space charge field in the dark
region and much lower field in the illuminated region. The applied
field is thus partially screened by the space charges induced by the
soliton forming optical beam.  The refractive index changes which is
proportional to this space charge field. The balance of self
diffraction of the optical beam by the focusing effect of the space
charge field induced nonlinearity leads to the formation of spatial
solitons. These screening solitons were first predicted by Segev et.
al.~\citep{47}, whereas, the experimental observation of bright SS
were reported by M.Shih et. al.~\citep{48} and those of dark SS were
reported by Z. Chen et. al.~\citep{49}.\\\\ The second type PR
soliton is the photovoltaic solitons ~\citep{57,58,59,60,61}, the
formation of which, however, requires an unbiased PR crystal that
exhibits photovoltaic effect, i.e., generation of dc current in a
medium illuminated by a light beam. The photovoltaic (PV) solitons
owe their existence to bulk photovoltaic effect, which creates the
space charge field, that, in turn modifies the refractive index and
gives rise to spatial solitons. These solitons were first predicted
by G. C. Valey et. al.~\citep{57} and observed experimentally in 1D
by M.Taya et.al.~\citep{58} and in 2D by Z.Chen et. al.~\citep{59}.
Two dimensional bright photovoltaic spatial solitons were also
observed in a Cu:KNSBN crystal by She et. al.~\citep{62}. The
observed spatial solitons were broader than those predicted by
considering $E_{sc}$ due to signal beam alone. This was
satisfactorily explained by an equivalent field induced by the
background field.\\\\ The third type of photorefractive solitons
arises when an electric field is applied to a photovoltaic
photorefractive crystal~\citep{31,60,61,63,64}. These solitons, owe
their existence to both photovoltaic effect and spatially nonuniform
screening of the applied field, and, are also known as screening
photovoltaic (SP) solitons. It has been verified that, if the bias
field is much stronger than the photovoltaic field, then, the
screening photovoltaic solitons are just like screening solitons. On
the other hand, if the applied field is absent, then SP solitons
degenerate into photovoltaic solitons in the closed circuit
condition. The first observation of bright photovoltaic screening
solitons in $LiNbO_{3}$  was reported by E. Fazio et.
al.~\citep{31}.

\subsection{Nonlinear Equation for Solitons}

In order to develop a semi analytical theory, one dimensional
reduction is introduced in the subsequent discussion. The optical
beam is such that no y dynamics is involved and it is permitted to
diffract only along $x$ direction. Electric field
$\overrightarrow{E}$ of the optical wave and the bias field $E_{0}$
are directed along the $x$ axis which is also the direction of the
crystalline c-axis. Under this special arrangement, the perturbed
extraordinary refractive index $\hat{n}_{e}$  is given
by~\citep{45,46,60}

\begin{equation}\label{e13}
({\hat{n}_{e})^{2}}=n_{e}^{2}-n_{e}^{4}r_{eff}E_{sc},
\end{equation}\
where $n_{e} $is the unperturbed extraordinary index of refraction.
This arrangement allows us to describe the optical beam propagation
 using Helmholtz's equation for the electric field
$\overrightarrow{E}$, which is as follows:

\begin{equation}\label{e14}
\nabla^{2}\overrightarrow{E}+(k_{0}\hat{n}_{e})^{2}\overrightarrow{E}=0,
\end{equation}

where $k_{0}=\frac{2\pi}{\lambda _{0}}$  and $\lambda_{0}$  is the
free space wavelength of the optical field. Moreover, if we assume
$\overrightarrow{E}= \overrightarrow{x}\Phi(x,z)exp[i(kz-\omega
t)],$
 where $k=k_{0}n_{e}$ and employ
slowly varying envelope approximation for the amplitude $\Phi $,
then, the Helmholtz's equation can be reduced to the following
parabolic equation:

\begin{equation}\label{e15}
i\frac{\partial \Phi}{\partial
z}+\frac{1}{2k}\frac{\partial^{2}\Phi}{\partial
x^{2}}-\frac{1}{2}k_{0}n_{e}^{3}r_{eff}E_{sc}\Phi=0.
\end{equation}\
The space charge field that has been evaluated earlier in section(5)
can be employed in the above equation to obtain following nonlinear
Schr\"{o}dinger equation:

\begin{equation}\label{e16}
i\frac{\partial A}{\partial\xi }+\frac{1}{2}\frac{\partial^{2}
A}{\partial s^{2}}-\beta(1+\rho)\frac{A}{1+|A|^{2}}-\alpha
\frac{(\rho-|A|^{2})A}{1+|A|^{2}}+\delta\frac{A}{1+|A|^{2}}\frac{\partial
|A|^{2}}{\partial s}=0,
\end{equation}\

where $\xi=\frac{z}{k_{0}n_{e}x_{0}^{2}}$, $s=\frac{x}{x_{0}}$,
$\rho=\frac{I_{\infty}}{I_{d}}$,
$\beta=(k_{0}x_{0})^{2}(n_{e}^{4}r_{eff}/2)E_{0}$,
$\alpha=(k_{0}x_{0})^{2}(n_{e}^{4}r_{eff}/2)E_{p}$,
$A=\sqrt{\frac{n_{e}}{2\eta_{0}I_{d}}}\Phi$   and
$\delta=(k_{0}^{2}x_{0}r_{eff}n_{e}^{4}k_{B}T)/(2e)$. Equation  (16)
gives rise to varieties of solitons depending on specific
experimental situations.  Important parameters which classify these
solitons and govern their dynamics are $\alpha $  and $\beta$. The
parameter $\delta$, which is associated with the diffusion term, is
not directly responsible for soliton formation. The diffusion
processes is primarily responsible for bending of trajectories of
propagating solitons, hence, large value of $\delta$ influences the
trajectory of bending. The dimensionless parameter $\beta$ can be
positive or negative depending on the sign of $E_{0}$ i.e., the
polarity of external applied field. In non-photovoltaic
photorefractive media $\alpha=0$,  therefore, if we neglect the
diffusion term, which is small indeed, and introduce only first
order correction, then $\beta$ is the parameter which governs the
soliton formation.    The relevant dynamical equation in
non-photovoltaic photorefractive media turns out to be
\begin{equation}\label{e17}
i\frac{\partial A}{\partial\xi }+\frac{1}{2}\frac{\partial^{2}
A}{\partial s^{2}}-\beta(1+\rho)\frac{A}{1+|A|^{2}}=0.
\end{equation}\

This is the  modified nonlinear Schr\"{o}dinger equation(MNLSE) with
saturable nonlinearity which is not integrable. The term $1/(1+
|A|^{2})$ represents local  saturable nonlinear change of refractive
index of the crystal induced by the optical beam. The saturating
nature of the nonlinearity will be more clearly evident if we make
the transformation $A=u\ exp [-i\beta(1+\rho)\xi]$, in which case
above equation reduces to

\begin{equation}\label{e18}
i\frac{\partial u}{\partial\xi }+\frac{1}{2}\frac{\partial^{2}
u}{\partial s^{2}}+\beta(1+\rho)\frac{|u|^{2}}{1+|u|^{2}}u=0.
\end{equation}\

It will be shown subsequently that a positive $\beta$ is essential
for creation of bright spatial solitons, whereas, a negative $\beta$
leads to the formation of  dark or gray solitons. Thus, by changing
the polarity of the external applied field, it is possible to create
bright as well as dark solitons in the same photorefractive media.
Equation(17) has been extensively investigated for bright, dark as
well as gray solitons~\citep{46,60,61}. In the next section, we
consider bright screening spatial solitons using a method outlined
by Christodoulides and Carvalho~\citep{46} and subsequently employed
in a large number of
investigations~\citep{108,109,110,111,137,145,146}.

\subsection{Bright Screening Spatial Solitons}

We fist consider  bright screening solitons for which the soliton
forming beam should vanish at infinity $(s\rightarrow\pm\infty)$,
and thus, $I_{\infty}=\rho=0$. Therefore, bright type solitary waves
should satisfy

\begin{equation}\label{e19}
i\frac{\partial A}{\partial\xi }+\frac{1}{2}\frac{\partial^{2}
A}{\partial s^{2}}-\beta\frac{A}{1+|A|^{2}}=0.
\end{equation}\

 We can obtain stationary bright solitary wave solutions by
expressing the beam envelope as
$A(s,\xi)=p^{\frac{1}{2}}y(s)\exp(i\nu \xi) $,  where $\nu$ is the
nonlinear shift in propagation constant and  y(s) is a normalized
real function which is bounded as, $0\leq y(s)\leq1.$ The quantity p
represents the ratio of the peak intensity $(I_{max})$  to the dark
irradiance $I_{d}$, where $I_{max}=I(s=0)$. Furthermore, for bright
solitons, we require $y(0)=1$, $\dot{y}(0)=0$ and
$y(s\rightarrow{\pm \infty})=0$ . Substitution of the ansatz for
$A(s,\xi)$ into equation(19) yields

\begin{equation}\label{e20}
\frac{d^{2}y}{ds^{2}}-2\nu y-2\beta\frac{y}{1+py^{2}}=0.
\end{equation}\

Integration of above equation once and use of boundary condition
yields

\begin{equation}\label{e21}
\nu=-(\beta/p)\ln(1+p),
\end{equation}\
and

\begin{equation}\label{e22}
\left(\frac{dy}{ds}\right)^{2}=(2\beta/p)[\ln(1+py^{2})-y^{2}\ln(1+p)].
\end{equation}\

By virtue of integration of above equation, we  immediately obtain

\begin{equation}\label{e23}
s=\pm\frac{1}{(2\beta)^{1/2}}\int_{y}^{1}\frac{p^{1/2}}{[\ln(1+p\widehat{y}^{2})-\widehat{y}^{2}\ln(1+p)]^{1/2}}
d\widehat{y}.
\end{equation}

The nature of above integrand  is such that it does not provide any
closed form solution. Nevertheless, the normalized bright profile
$y(s)$ can be easily obtained by the use of simple numerical
procedure. It  can be easily shown that the quantity in the square
bracket in equation(23) is always positive for all values of $y^{2}$
between  $0\leq y(s)\leq 1$.  Therefore, the bright screening
photorefractive solitons can exist in a medium only when $\beta>0 $
i.e., $E_{0}$ is positive. For a given value of  $\beta$, the
functional form of $y $ can be obtained for different p which
determines  soliton profile. For a given physical system, the
spatial beam width of these solitons depends on two parameters
$E_{0}$  and $p $. For illustration, we take SBN crystal with
following parameters $n_{e}=2.35$, $ r_{33}=224\times10^{-12}$ m/V.
Operating wavelength $\lambda_{0}=0.5\mu m $, $x_{0}=20\mu m$    and
$E_{0}=2\times10^{5}$ V/m. With these parameters, value of
$\beta\simeq 43$.  Figure (2) depicts typical normalized intensity
profiles of bright solitons.
\begin{center}
\textit{Insert Figure (2) here}
\end{center}

\subsubsection{Bistable Screening  Solitons}

Equation(19) possesses several conserved quantities, one such
quantity is $P=\int_{\infty}^{\infty}|A|^{2} ds$, which can be
identified as the total power of the soliton forming optical beam.
Numerically evaluated soliton profile $|A|^{2}=p|y(s)|^{2}$ can be
employed to calculate $P$. Solitons obtained from equation(23) are
stable since they obey Vakhitiov and Kolokolov~\citep{65} stability
criteria i.e., $dP/d\nu>0$. These soliton profiles can be also
employed to find out spatial width $\tau_{FWHM}$ ( full width at
half maximum) of solitons. Figure (3) demonstrates the variation of
spatial width $\tau_{FWHM}$ with  $p$. This figure signifies the
existence of two-state solitons, also known as bistable solitons,
i.e., two solitons possessing same spatial width   but
 different power. Similar bistable solitons were earlier predicted
in doped fibers~\citep{66,67}. However, these bistable solitons are
different from those which were predicted by Kaplan and
others~\citep{68}, where two solitons with same power possessing two
different nonlinear propagation constant.
\begin{center}
\textit{Insert Figure (3) here}
\end{center}

Another point worth mentioning is that, the $\tau_{FWHM}$ vs $ p$
curve in figure(3) possesses local minimum, hence, it is evident
that these solitons can exist only if their spatial width is above
certain minimum value.  This minimum value increases with the
decrease in the value of $\beta$. For illustration, the shapes of a
typical pair of bistable solitons with same $\tau_{FWHM}$  but
different peak power have been depicted in figure(4). The dynamical
behavior of these bistable solitons, while propagating, can be
examined by full numerical simulation of equation(19), which
confirms their stability.

\begin{center}
\textit{Insert Figure (4) here}
\end{center}

\subsection{Dark Screening Solitons}

Equation (17) also yields dark solitary wave
solutions~\citep{46,69}, which
 exhibit anisotropic field profiles with respect
to $s$. These solitons are embedded in a constant intensity
background, therefore,  $I_{\infty}$ is finite, hence, $\rho$ is
also finite. To obtain stationary solutions, we assume
$A(s,\xi)=\rho^{1/2} y(s)exp(i\nu\xi)$, where, like earlier case,
$\nu$ is the nonlinear shift in propagation constant and  $y(s)$ is
a  normalized real odd function of $s$. The profile $y(s)$ should
satisfy following properties: $y(0)=0, y(s\rightarrow\pm
\infty)=\pm1, \frac{dy}{ds}=\frac{d^{2}y}{ds^{2}}=0$  as
$s\rightarrow \pm\infty$.\ Substituting the form of A in
equation(17) we obtain,

\begin{equation}\label{e24}
\frac{d^{2}y}{ds^{2}}-2\nu y-2\beta ( 1+\rho)\frac{y}{1+\rho
y^{2}}=0.
\end{equation}\

Integrating above equation once and employing boundary condition, we
get

\begin{equation}\label{e25}
\nu=-\beta.
\end{equation}\

Following similar procedure as employed earlier, we immediately
obtain

\begin{equation}\label{e26}
s=\pm\frac{1}{(-2\beta)^{1/2}}\int_{y}^{0}\frac{d\widehat{y}}{[(\widehat{y}^{2}-1)-\frac{(1+\rho)}{\rho}\ln\frac{1+\rho
\widehat{y}^{2}}{1+\rho}]^{1/2}}.
\end{equation}\

The quantity within the square bracket in equation(26) is always
positive for all values of $y^{2}\leq 1$, thus $\beta$ i.e., $E_{0}$
must be negative so that r.h.s of equation(26) is not imaginary. An
important point to note is that, for a particular type of
photorefractive material, for example SBN, if positive polarity of
$E_{0}$ is required for bright solitons, then one can observe dark
solitons by changing  polarity of $E_{0}$ . Spatial width of these
solitons depends on only two variables $\beta$ and $\rho$ i.e.,
$E_{0}$  and $I_{\infty}$. It has been confirmed that, unlike their
bright counterpart, dark screening photovoltaic solitons do not
possess bistable property. For illustration,  we take the same SBN
crystal with other parameters unchanged, except in the present case
$E_{0}=-2\times10^{5}$ V/m. Therefore $\beta\approx-43$. Figure (5)
depicts normalized intensity profiles of dark solitons which are
numerically identified using equation(26).

\begin{center}
\textit{Insert Figure (5) here}
\end{center}

\subsection{Gray Screening Solitons   }\ \
Besides bright and dark solitons, equation (17) also admits another
interesting class of solitary waves, which are known as gray
solitons~\citep{46}. In this case too, wave power density attains a
constant value at infinity i.e., $I_{\infty}$  is finite, and hence,
$\rho$ is finite. To obtain stationary solutions, we assume

\begin{equation}\label{e27}
A(s,\xi)=\rho^{1/2}y(s)\exp[i(\nu\xi+\int^{s}\frac{Jd\widehat{s}}{y^{2}(\widehat{s})})],
\end{equation}\

where $\nu$ is again the nonlinear shift in propagation constant,
$y(s)$ is a  normalized real even function of $s$   and $J$  is a
real constant to be determined. The normalized real function
satisfies the boundary conditions $y^{2}(0)=m(0<m<1)$, i.e., the
intensity is finite at the origin,  $\dot{y(0)}=0$, $y(s\rightarrow
\pm\infty)=1$ and all derivatives of $y(s)$  are zero at infinity.
The parameter  m describes grayness, i.e., the intensity  $I(0)$ at
the beam center  is $I(0)=mI_{\infty}$.  Substitution of the above
ansatz for A in equation(17) yeilds
\begin{equation}\label{e28}
\frac{d^{2}y}{ds^{2}}-2\nu y-2\beta ( 1+\rho)\frac{y}{1+\rho
y^{2}}-\frac{J^{2}}{y^{3}}=0.
\end{equation}\

Employing boundary conditions on y at infinity we obtain
\begin{equation}\label{e29}
J^{2}=-2(\nu+\beta).
\end{equation}

Integrating equation(28) once and employing appropriate boundary
condition, we immediately obtain

\begin{equation}\label{e30}
\nu=\frac{-\beta}{(m-1)^{2}}\left[\frac{m(1+\rho)}{\rho}\ln\left(\frac{1+\rho
m}{1+\rho}\right)+1-m\right].
\end{equation}

Finally

\begin{equation}\label{e31}
\left(\frac{dy}{ds}\right)^{2}=2\nu(y^{2}-1)+\frac{2\beta}{\rho}(1+\rho)\ln
\left(\frac{1+\rho y^{2}}{1 + \rho
}\right)+2(\nu+\beta)\left(\frac{1-y^{2}}{y^{2}}\right).
\end{equation}\

The normalized amplitude $y(s)$ can be obtained by numerical
integration of above equation. Note that dark solitons are a
generalization of these gray solitons. Unlike bright or dark
solitons, the phase of gray solitons is not constant across $s$,
instead varies across $s$.  Existence of these solitary waves are
possible only when $\beta<1$  and   $ m<1$.

\subsection{Self Deflection of Bright Screening Solitons  }

In the foregoing discussion we have neglected diffusion, however,
the effect of diffusion cannot be neglected when solitons spatial
width is comparable with the diffusion length. The diffusion process
introduces asymmetric contribution in the refractive index change
which causes solitons to deflect during propagation. Results of
large number of investigations  addressing the deflection of
photorefractive spatial solitons in both non-centrosymmetric and
centrosymmetric photorefractive crystals are now
available~\citep{70,71,72,73,74,75,76,77,78,79,80,81,82}. Several
authors have investigated self bending phenomenon using perturbative
procedure~\citep{83,84}. In this section, we employ a method of
nonlinear optics~\citep{85,86}  which is different from perturbative
approach. To begin with, we take finite $\delta$ and use the
following equation to study the self bending of screening bright
spatial solitons

\begin{equation}\label{e32}
i\frac{\partial A}{\partial\xi }+\frac{1}{2}\frac{\partial^{2}
A}{\partial
s^{2}}-\beta\frac{A}{(1+|A|^{2})}+\delta\frac{A}{(1+|A|^{2})}\frac{\partial
(|A|^{2})}{\partial s}=0.
\end{equation}

To obtain stationary solitary waves, we make use of the ansatz

\begin{equation}\label{e33}
A(s,\xi)=A_{0}(s,\xi)\exp[-i\Omega(s,\xi)],
\end{equation}

in equation(32).   A straightforward calculation yields following
equations:

\begin{equation}\label{e34}
\frac{\partial A_{0}}{\partial\xi}-\frac{\partial A_{0}}{\partial
s}\frac{\partial\Omega}{\partial
s}-\frac{1}{2}A_{0}\frac{\partial^{2}\Omega}{\partial s^{2}}=0,
\end{equation}\
and

\begin{equation}\label{e35}
A_{0}\frac{\partial
\Omega}{\partial\xi}+\frac{1}{2}\frac{\partial^{2} A_{0}}{\partial
s^{2}}-\frac{1}{2}A_{0}\left(\frac{\partial\Omega}{\partial
s}\right)^{2}-\beta\frac{A_{0}}{1+A_{0}^{2}}+\frac{\delta}{1+A_{0}^{2}}\frac{\partial(A_{0}^{2})}{\partial
s}=0.
\end{equation}\\
We look for a self-similar solution of (34) and (35) of the form

\begin{equation}\label{e36}
A_{0}=\frac{A_{00}}{\sqrt{f(\xi)}}exp\left[-\frac{({s-s_{0}(\xi)})^{2}}{2r_{0}^{2}f^{2}(\xi)}\right],
\end{equation}

\begin{equation}\label{e37}
\Omega(s,\xi)=\frac{(s-s_{0}(\xi))^{2}}{2}\Lambda_{1}(\xi)+(s-s_{0}(\xi))\Lambda_{2}(\xi)+\Lambda_{3}(\xi),
\end{equation}

\begin{equation}\label{e38}
\Lambda_{1}=-\frac{1}{f(\xi)}\frac{df}{d\xi}, \ \ \ \ \ and \ \ \ \
\ \Lambda_{2}=-\frac{ds_{0}}{d\xi},
\end{equation}

where,  $r_{0}$  is a constant and   $f(\xi)$   is a parameter which
together with   $r_{0}$ describe spatial width; in particular,
$r_{0}f(\xi)$ is  the spatial  width of the soliton and
$\Lambda_{3}$ is an arbitrary longitudinal phase function. $s_{0}
(\xi)$ is the location of the center of the soliton. For a
nondiverging/ nonconverging soliton,  $f(\xi)=1$. Moreover, we
assume that initially solitons are nondiverging i.e.,
$\frac{df(\xi)}{d\xi}=0$ at $\xi=0$. Substituting for $A_{0}$ and
$\Omega$ in equation(35) and equating coefficients of $s$    and
$s^{2}$ on both sides, we obtain

\begin{equation}\label{e39}
\frac{d^{2}f}{d\xi^{2}}=\frac{1}{r_{0}^{4}f^{3}}-\beta\frac{2A_{00}^{2}}{r_{0}^{2}f^{2}}\left(1+\frac{A_{00}^{2}}{f}\right)^{-2},
\end{equation}
and

\begin{equation}\label{e40}
\frac{d^{2}s_{0}(\xi)}{d\xi^{2}}=-\delta\frac{2A_{00}^{2}}{r_{0}^{2}f^{3}}\left(1+\frac{A_{00}^{2}}{f}\right)^{-1}.
\end{equation}\

Equation(39) describes the dynamics of the width of soliton  as it
propagates in the medium, while equation(40) governs the dynamics of
the centre of the soliton. In order to find out how a stationary
soliton deviates from its initial propagation direction, we first
solve equation(39) for stationary soliton states. From equation(39),
condition for stationary soliton states can be obtained as

\begin{equation}\label{e41}
r_{0}=\left[\frac{(1+A_{00}^{2})^{2}}{2\beta
A_{00}^{2}}\right]^{1/2}.
\end{equation}\

\begin{center}
\textit{Insert Figure (6) here}
\end{center}

Figure (6) depicts variation of solitons spatial width with the
normalized peak power  $A_{00}^{2}$  .  The curve in figure (6) is
the existence curve of stationary solitons. Each point on this curve
signifies the existence of a stationary soliton with a given spatial
width and peak power.  A stationary soliton of specific power and
width as described by equation(41) deviates from its initial path
which can be found out by integrating equation(40).  The  equation
of trajectory of the center of  soliton is

\begin{equation}\label{e42}
s_{0}(\xi)=-\frac{\Theta}{2}\xi^{2}+s_{00},
\end{equation}\

where $ \Theta=\delta\frac{2A_{00}^{2}}{r_{0}^{2}}(1+A_{00}^{2})$,
$s_{00}=s_{0}(\xi=0)$, moreover $ f=1 $, since we are only
interested in stationary solitons whose spatial shape remain
invariant.  Thus, the beam center follows a parabolic trajectory.
The displacement of the soliton center with propagation distance has
been depicted in figure (7). It is evident that the peak power of
soliton influences lateral displacement of the soliton centre. The
lateral displacement suffered  by  the spatial soliton is given by
$\frac{\Theta}{2}\xi^{2}$. Equation(42) implies that the angular
displacement of the soliton center shifts linearly with the
propagation distance $\xi$. The more explicit expression of the
angular displacement, i.e., the angle between the center of the
soliton and the z axis can be easily obtained as  $\Theta\xi$.

\begin{center}
\textit{Insert Figure (7) here}
\end{center}
\subsection{Photovoltaic and Screening Photovoltaic Solitons  }

 Steady state photovoltaic solitons  can be created  in a photovoltaic photorefractive crystal
 without a bias field. These photovoltaic solitons result from the photovoltaic effect~\citep{57,58,59,60,61} .
  Recently, it has been predicted theoretically that the screening-photovoltaic(SP) solitons are
  observable  in the steady state when an external electric field is applied to a photovoltaic photorefractive crystal~\citep{31,78}.
   These SP solitons result from both the photovoltaic effect and spatially nonuniform screening of the applied field.

    If the bias field is much stronger in comparison to  the photovoltaic field, then the SP solitons are just like
    the screening solitons. In absence of the  applied field, the SP solitons degenerate into
    photovoltaic solitons in the close-circuit condition. In other words, a closed-circuit
    photovoltaic soliton or screening soliton is a special case of the SP soliton. Thus,
     in the subsequent analysis, we develop theory for SP solitons and obtain solutions
     of photovoltaic solitons as a special case.
 When the diffusion process is ignored, the dynamics
  of these steady-state screening PV solitons
  ( bright, dark and gray) can be examined~\citep{60,78,88}
   using the following equation

\begin{equation}\label{e43}
i\frac{\partial A}{\partial\xi }+\frac{1}{2}\frac{\partial^{2}
A}{\partial
s^{2}}-\beta(1+\rho)\frac{A}{(1+|A|^{2})}-\alpha\frac{(\rho-|A|^{2})A}{(1+|A|^{2})}=0.
\end{equation}\

Adopting the procedure, employed earlier, the bright soliton profile
of screening photovoltaic solitons~\citep{60}  turns out to be

\begin{equation}\label{e44}
s(2(\alpha+\beta))^{1/2}=\pm\int_{y}^{1}\frac{p^{1/2}}{[\ln(1+p\widehat{y}^{2})-\widehat{y}^{2}\ln(1+p)]^{1/2}}
d\widehat{y},
\end{equation}\

The  quantity within the bracket in the right hand side  is always
positive for $y(s)\leq 1$, thus, $(\alpha+\beta)$  must be positive
for bright screening photovoltaic solitons. From equation(44), it is
evident that, these bright SP solitons result both from the
photovoltaic effect $(\alpha\neq 0)$ and from spatially nonuniform
screening $(\beta\neq 0)$ of the applied electric field in a biased
photovoltaic-photorefractive crystal. Formation of these solitons
depends not only on the external bias field but also on the
photovoltaic field. When we set $\alpha =0$, these solitons are just
like screening solitons in a biased nonphotovoltaic photorefractive
crystal~\citep{46}. In addition, when $\beta=0$ ,  we obtain
expression of bright photovoltaic solitons in the close circuit
realization~\citep{61}. Thus, these SP solitons differ both from
screening solitons in a biased nonphotovoltaic photorefractive
crystal and from photovoltaic solitons in a photovoltaic
photorefractive crystal without an external bias field. One
important point to note is that the experimental conditions are
different for creation of screening, PV and SP solitons. SP solitons
can be created  in a biased photovoltaic-photorefractive crystal,
whereas, creation of screening solitons are possible in biased
nonphotovoltaic-photorefractive crystals. The PV solitons can be
created  in photovoltaic-photorefractive crystal without an external
bias field.\\\\\ We can also derive dark solitons from  equation
(43), the normalized dark field profile can be easily obtained as

\begin{equation}\label{e45}
[-2(\alpha+\beta)]^{1/2}s=\pm\int_{y}^{0}\frac{d\widehat{y}}{\left[(\widehat{y}^{2}-1)-\frac{1+
\rho}{\rho}\ln(\frac{1+ \rho\widehat{y}^{2}}{1+\rho})\right]^{1/2}},
\end{equation}\

The condition for existence of dark solitons is $(\alpha+\beta)<0$.
In a medium like $LiNbO_{3}$, $\alpha<0$, therefore,   if
$|\beta|<|\alpha|$ then the dark SP solitons  can be observed
irrespective of the polarity of external bias field. However,
photovoltaic constant in some photovoltaic materials, such as,
$BaTiO_{3}$ depends on polarization of light~\citep{87}. This means
sign of $\alpha$ may be positive or negative depending on
polarization of light. Therefore, to observe dark SP solitons,
polarization of light and external bias must be appropriate so that
$(\alpha+\beta)<0$ . It is evident from the expression of dark
solitons that, if bias field is much stronger in comparison to the
photovoltaic field, then these SP dark solitons are just like
screening dark solitons. On the other hand, if the applied external
field is absent, then these dark SP solitons degenerate into
photovoltaic dark solitons in the closed circuit condition.

\subsection{Self Deflection of   Photovoltaic and Screening Photovoltaic
Solitons  }

 In absence of diffusion process, which is   usually weak,
 photovoltaic solitons propagate along a straight line keeping their shape
  unchanged. However, when the spatial width
  of   soliton is small, the diffusion effect is significant, which
  introduces an asymmetric tilt in the light-induced photorefractive
  waveguide,  that in turn is expected to affect the propagation
  characteristics of  steady-state photorefractive solitons.
  Several authors~\citep{74,78,79,80,81,82,88}  have examined the self bending
  phenomenon of   PV solitons.  The effects of higher order space
  charge field on this self bending phenomenon have been also
  investigated~\citep{81,82,88}.\\\
The deflection of PV bright solitons depends on the strength of
photovoltaic field. When the PV field $E_{p}$ is less than  a
certain characteristic value, solitons  bend opposite to crystal
c-axis and absolute value of spatial shift due to first order
diffusion is always larger in comparison to that due to both first
and higher order diffusion~\citep{79}. When  $E_{p}$ is larger than
the characteristic value, direction of bending depends on the
strength of $E_{p}$ and the input intensity of soliton forming
optical beam. Self deflection can be completely arrested  by
appropriately selecting  $E_{p}$ and intensity of the  soliton
forming optical beam.\\\\\ PV dark solitons experience approximately
adiabatic self deflection in the direction of the c-axis of the
crystal and the spatial shift follows an approximately parabolic
trajectory. Nature of self deflection of PV dark solitons is
different from that of bright solitons in which self deflection
occurs in the direction opposite to the c-axis of the crystal.
Effect of higher order space charge field on self deflection has
been also investigated~\citep{88}, which indicates a considerable
increase in the self deflection of dark solitons, especially under
the high PV field. Thus, the spatial shift due to both first and
higher order diffusion is larger in
comparison to that when first order diffusion is present alone.\\\\
The self deflection of screening PV solitons depends on both the
bias and PV fields~\citep{81}. When the bias field is positive and
the PV field is negative, the screening PV bright solitons always
bend in the direction opposite to the crystal c-axis, and the
absolute value of the spatial shift
 due to  first order diffusion term alone is always
smaller than that  due to both first  and higher order. When PV
field is positive and bias field is negative or both are positive,
then the bending direction depends both on the strength of two
fields and on the intensity of the optical beam. Bending can be
completely compensated for appropriate polarity of the two fields
and optical beam  intensity.
\section{Two-Photon Photorefractive Phenomenon }

Three types of steady state photorefractive spatial solitons, as
elucidated earlier, owe their existence on the single photon
photorefractive phenomenon.   Recently, a new kind of
photorefractive solitons has been proposed in which the soliton
formation mechanism relies on two-photon photorefractive phenomenon.
 It is understood that the
two-photon process can significantly enhance the photorefractive
phenomenon. A new model has been introduced by Castro-Camus and
Magana~\citep{91}  to investigate two-photon photorefractive
phenomenon. This model includes a valance band (VB), a conduction
band(CB) and an intermediate allowed level(IL). A gating beam with
photon energy $\hbar\omega_{1}$ is used to maintain a fixed quantity
of excited electrons from the valance band  to the intermediate
level, which are then excited to the conduction band by the signal
beam with photon energy $\hbar\omega_{2}$. The signal beam induces a
charge distribution that is identical to its intensity distribution,
which in turn gives rise to a nonlinear change of refractive index
through space charge field. Very recently, based on Castro-Camus and
Magana's model, Hou et. al., predicted that two-photon screening
solitons (TPSS)  can be created in a biased nonphotovoltaic
photorefractive crystal~\citep{92}  and two-photon photovoltaic
(TPPV) solitons can  be also created in a PV crystal under
open-circuit condition~\citep{93}. Recently, the effect of external
electric field on screening photovoltaic solitons due to two-photon
photorefractive phenomenon has been also investigated
~\citep{94,95}.

\subsection{Two-Photon Photorefractive Nonlinearity }

In order to estimate the optical nonlinearity arising out in two
photon-photorefractive media, we consider an  optical configuration
whose schematic diagram is shown in figure (8).

\begin{center}
\textit{Insert Figure (8) here}
\end{center}

The electrical circuit consist of  a crystal ( could be made of
photovoltaic-photorefractive  or non photovoltaic-photorefractive
material), external electric field bias voltage V and external
resistance   $R$. $V_{0}$   and  $E_{0}$ respectively denote the
potential and electric field strength between the crystal electrodes
which are separated by a distance d. Assuming  the spatial extent of
optical wave is much less than d, we
have     $E_{0}=V_{0}/d$;        additionally $V/d=E_{a}$.\\\

Therefore,
\begin{equation}\label{e46}
V=E_{0}d+JSR=E_{a}d,
\end{equation}\

where  $S$ is the surface area of the crystal electrodes and $J$ is
the current density. The soliton forming optical beam with intensity
$I_{2}$ propagates along the $z$ direction of the crystal and is
permitted to diffract only along the $x$ direction. The optical beam
is polarized along the $x$ axis which is also the direction of
crystal c-axis and the external bias field is also directed along
this direction. The crystal is illuminated with a gating beam of
constant intensity $I_{1}$. The space charge field $E_{sc}$ due to
two-photon photorefractive phenomenon can be obtained from the set
of rate, current, and Poisson's equations proposed by Castro-Camus
et. al.~\citep{91}.  In the steady state, these equations are

\begin{equation}\label{e47}
(s_{1}I_{1}+\beta_{1})(N_{D}-N_{D}^{+})-\gamma_{1}N_{i}N_{D}^{+}-\gamma_{R}NN_{D}^{+}=0,
\end{equation}

\begin{equation}\label{e48}
(s_{1}I_{1}+\beta_{1})(N_{D}-N_{D}^{+})+\gamma_{2}N(N_{it}-N_{i})-\gamma_{1}N_{i}N_{D}^{+}-(s_{2}I_{2}+\beta_{2})N_{i}=0,
\end{equation}

\begin{equation}\label{e49}
(s_{2}I_{2}+\beta_{2})N_{i}+\frac{1}{e}\frac{\partial J}{\partial
x}-\gamma_{R}NN_{D}^{+}-\gamma_{2}N(N_{it}-N_{i})=0,
\end{equation}

\begin{equation}\label{e50}
\epsilon_{0}\epsilon_{r}\frac{\partial E_{sc}}{\partial
x}=e(N_{d}^{+}-N-N_{i}-N_{A}) ,
\end{equation}

\begin{equation}\label{e51}
J=e\mu N
E_{sc}+\kappa_{p}s_{2}(N_{D}-N_{D}^{+})I_{2}+eD\frac{\partial
N}{\partial x},
\end{equation}

\begin{equation}\label{e52}
\frac{\partial J}{\partial x}=0, \ \ \ \\\\\\\\ J=constant,
\end{equation}

where  $N_{D}$ ,  $N_{D}^{+}$ ,   $N_{A}$   and     $N$    are the
donor density, ionized donor density,      acceptor or trap density
and  density of electrons in the conduction band, respectively.
$N_{i}$ is the density of electrons in the intermediate state,
$N_{it}$  is the density of traps in the intermediate state.
$\kappa_{p}$ , $\mu$ and $e$  are respectively the photovoltaic
constant, electron mobility and electronic charge; $\gamma_{R}$ is
the recombination factor of the conduction to valence band
transition, $\gamma_{1}$ is the recombination factor for
intermediate allowed level to valence band transition, $\gamma_{2}$
is the recombination factor of the conduction band to intermediate
level transition; $\beta_{1}$ and $\beta_{2}$ are respectively the
thermo-ionization probability constant for transitions from valence
band to intermediate level and intermediate level to conduction
band; $s_{1}$ and $s_{2}$ are photo-excitation crosses. $D$  is the
diffusion coefficient and  $I_{2}$ is the intensity of the soliton
forming beam. We adopt the usual approximations $N_{D}^{+}\sim
N_{A}$   and $N_{it}-N_{i}<<N_{A}$ . In addition, we also assume
that, the power density  is uniform at large distance from the
center of the soliton forming beam, thus, at $x\rightarrow
\pm\infty,\ I_{2} (x\rightarrow \pm\infty, z)=
constant=I_{2\infty}$. Obviously, the space charge field in the
above remote region is also uniform, i.e., $E_{sc} (x\rightarrow
\pm\infty, z)=E_{0}$. The space charge field can be obtained using
standard procedure~\citep{94}, which turns out to be

\begin{eqnarray}\label{e53}
E_{sc}&=&gE_{a}\frac{(I_{2\infty
}+I_{2d})(I_{2}+I_{2d}+\frac{\gamma_{1}N_{A}}{s_{2}})}{(I_{2
}+I_{2d})(I_{2\infty}+I_{2d}+\frac{\gamma_{1}N_{A}}{s_{2}})}\nonumber \\
& & { }+E_{p}\frac{s_{2}(gI_{2\infty
}-I_{2})(I_{2}+I_{2d}+\frac{\gamma_{1}N_{A}}{s_{2}})}{(I_{2
}+I_{2d})(s_{1}I_{1}+\beta_{1})}\nonumber  \\
& &{ }-\frac{D}{\mu s_{2}}\frac{\gamma_{1}N_{A}}{(I_{2
}+I_{2d})(I_{2\infty}+I_{2d}+\frac{\gamma_{1}N_{A}}{s_{2}})}\frac{\partial
I_{2}}{\partial x},
\end{eqnarray}

where   $E_{p}=\kappa_{p} N_{A}\gamma_{R}/e\mu $      is the
photovoltaic field, $I_{2d}=\beta_{2}/s_{2}$  is the so called dark
irradiance. $g= 1/(1+q)$, $q= \frac{e\mu N_{\infty}SR}{d}$ ,
$N_{\infty}=N( x\rightarrow \pm\infty )$. In general the parameter
$g$ is a positive quantity and is bounded between $0\leq g\leq 1$.
Under short circuit condition  $R=0$ and $g=1$, implying that the
external electric field is totally applied to the crystal. For open
circuit condition $R\rightarrow \infty,$  thus,   $g=0$ i.e., no
bias field is applied to the crystal.

\subsection{Nonlinear Evolution Equation  }\ \ \ \ \ \ \ \ \ \ \ \ \
As usual, the optical field   of the incident soliton forming beam
 is taken as
$\overrightarrow{E}=\overrightarrow{x}\Phi(x,z)\exp[i(kz-\omega
t)]$, where $k=k_{0} n_{e},  k_{0}=2\pi /\lambda_{0} $,
$\lambda_{0}$ is the free space wavelength of the optical field,
$n_{e}$ is the unperturbed extraordinary index of refraction and
$\Phi$ is the slowly varying envelope of the optical field.
Employing the space charge field as given by equation(53) and
following the procedure employed earlier, the nonlinear
Schr\"{o}dinger equation for the normalized envelope $A(s,\xi)$ can
be obtained as~\citep{94}
\begin{eqnarray}\label{e54}
i\frac{\partial A}{\partial\xi }+\frac{1}{2}\frac{\partial^{2}
A}{\partial s^{2}}-\beta
g\frac{(1+\rho)(1+\sigma+|A|^{2})A}{(1+|A|^{2})(1+\sigma+\rho)}-& &
\alpha\eta\frac{(g\rho-|A|^{2})(1+\sigma+|A|^{2})A}{1+|A|^{2}}\nonumber\\
& & { }+\frac{\delta\sigma
A}{(1+|A|^{2})(1+\sigma+|A|^{2})}\frac{\partial |A|^{2}}{\partial
s}=0,
\end{eqnarray}\ \ \ \

where   $\rho= I_{2\infty}/I_{2d}$,    $A=
\sqrt{\frac{n_{e}}{2\eta_{0}I_{2d}}}\Phi$, $\xi=z/(k_{0} n_{e}
x_{0}^{2})$, $s=x/x_{0}$,    $\beta=(k_{0} x_{0} )^{2} (n_{e}^{4}
r_{33}/2) E_{a}$, $\alpha=(k_{0} x_{0})^{2} (n_{e}^4 r_{33}/2)
E_{p}$, $\delta=(k_{0} x_{0} )^2 (n_{e}^{4} r_{33}/2)D/(x_{0}\mu)$,
$\eta=\beta_{2}/(s_{1} I_{1}+\beta_{1}),
\sigma=\frac{\gamma_{1}N_{A}}{s_{2}I_{2d}}=\frac{\gamma_{1}N_{A}}{\beta_{2}}$
and $\eta_{0}=\sqrt{\mu_{0}/\epsilon_{0}}. \ r_{33}$ is the
electro-optic coefficient of the two-photon photorefractive crystal.
Equation(54) can be employed to investigate screening, photovoltaic
and screening photovoltaic solitons under appropriate experimental
configuration.

\subsection{Screening Solitons }

For screening solitons, the crystal should be
nonphotovoltaic-photorefractive(i.e.,\ $\alpha=0$). Assuming that
the external bias field is totally applied to the crystal{(R=0)},
thus, $E_{a}=E_{0}$ and  $g=1$. Neglecting diffusion, the expression
for space charge field $E_{sc}$ turns out to be~\citep{91,92}

\begin{equation}\label{e55}
E_{sc}=E_{0}\frac{(I_{2\infty
}+I_{2d})(I_{2}+I_{2d}+\frac{\gamma_{1}N_{A}}{s_{2}})}{(I_{2
}+I_{2d})(I_{2\infty}+I_{2d}+\frac{\gamma_{1}N_{A}}{s_{2}})}.
\end{equation}
Note that though the gating beam of constant intensity $I_{1}$    is
required to maintain a quantity of excited electrons density in the
intermediate allowed level, it does not appear in the expression of
space charge field. The relevant modified nonlinear Schr\"{o}dinger
equation is obtained as

\begin{equation}\label{e56}
i\frac{\partial A}{\partial\xi }+\frac{1}{2}\frac{\partial^{2}
A}{\partial s^{2}}-\beta
\frac{(1+\rho)(1+\sigma+|A|^{2})A}{(1+|A|^{2})(1+\sigma+\rho)}=0.
\end{equation}

Fundamental properties of  dark, bright and grey  solitary wave
solutions of this equation have been investigated extensively by
several authors~\citep{92,96,97,98}.  Self deflection of these
solitons due to diffusion~\citep{92} and effects of higher order
space charge field on self deflection have been also
examined~\citep{97}. Jiang et. al.~\citep{98}  have examined
temporal behavior of these solitons. They predicted that, in the low
amplitude regime, FWHM of solitons will decrease monotonically to a
minimum steady state value,  and that the transition time of such
solitons should be independent of $\beta$ or soliton intensity and
is close to $10T_{d}$, where $T_{d}$ is the dielectric relaxation
time. They also predicted that the temporal properties of dark
solitons are similar to those of the bright solitons.\\\\\
Intensity of bright solitons vanishes at infinity, thus,
$I_{2\infty}=\rho=0$. Therefore,

\begin{equation}\label{e57}
i\frac{\partial A}{\partial\xi }+\frac{1}{2}\frac{\partial^{2}
A}{\partial
s^{2}}-\frac{\beta}{1+\sigma}\left(1+\frac{\sigma}{1+|A|^{2}}\right)A
=0.
\end{equation}\

Assume a bright soliton of the form $A=p^{1/2} y(s)\exp(i\nu\xi)$,
where $\nu$ is the nonlinear shift in propagation constant and
$y(s)$ is a  normalized real function, which is bounded as $0\leq
y(s)\leq 1$. The parameter  $p$  stands for the ratio of the peak
 intensity of the soliton to  dark irradiance $I_{2d}$. The
profile of the soliton turns~\citep{92}    out to be

\begin{equation}\label{e58}
s=\pm\int_{y}^{1}\frac{[\frac{2\beta\sigma}{(1+\sigma)}]^{-1/2}p^{1/2}}{[\ln(1+p\widehat{y}^{2})-\widehat{y}^{2}\ln(1+p)]^{1/2}}
d\widehat{y},
\end{equation}
while the expression for nonlinear phase shift is given by

\begin{equation}\label{e60}
\nu=-\left(\frac{\beta}{1+\sigma}\right)[1+\frac{\sigma}{p}\ln(1+p)].
\end{equation}

From equation(58), it can be easily shown that the bright soliton
requires $\beta>0$ i.e.,  $E_{0}>0$.  Therefore, screening bright
spatial solitons can be formed in a two-photon photorefractive
medium only when external bias field is applied in the same
direction with respect to the optical c-axis. FWHM of these spatial
solitons is inversely proportional to the square root of the
absolute value of the external bias field. In the low amplitude
limit  i.e.,when  $|A|^{2}<<1$, equation(57)  reduces to

\begin{equation}\label{e60}
i\frac{\partial A}{\partial\xi }+\frac{1}{2}\frac{\partial^{2}
A}{\partial
s^{2}}-\frac{\beta}{1+\sigma}\left(1+\sigma-\sigma|A|^{2}\right)A
=0.
\end{equation}\\
Above equation can be exactly solved analytically and the
one-soliton solution is

\begin{equation}\label{e61}
A(s,\xi)=p^{1/2}sech\left[\left(\frac{\beta p
\sigma}{1+\sigma}\right)^{1/2}s\right]\\exp\left[i\frac{\beta(p\sigma-2\sigma-2)}{2(1+\sigma)}\xi\right].
\end{equation}\\
The spatial width ($\tau_{FWHM}$) of these solitons are
$\tau_{FWHM}=2\ln(1+\sqrt{2})
\left(\frac{1+\sigma}{p\sigma\beta}\right)^{1/2}$. \\\\

 To obtain dark soliton solution, we take
$A(s,\xi)=\rho^{1/2}y(s)\exp[i\mu\xi]$,    where $y(s)$ is a
normalized odd function of s and satisfies the following boundary
conditions: $y(s=0)=0$, $y(s\rightarrow \pm \infty)=±1$,  and all
the derivatives of $y(s)$ vanish at infinity. The profile $y(s)$ of
these solitons can be obtained~\citep{92} using the following
relationship

\begin{equation}\label{e62}
s=\pm\int_{y}^{0}\frac{[\frac{-2\beta\sigma}{(1+\sigma+\rho)}]^{-1/2}}{\left[(\widehat{y}^{2}-1)-\frac{(1+\rho)}{\rho}\ln\left(\frac{1+\rho\widehat{y}^{2}}{1+\rho}\right)\right]^{1/2}}
d\widehat{y},
\end{equation}\

and the nonlinear phase shift  $\mu=-\beta$.  The dark solitons
require $\beta < 0$. In the low amplitude limit  i.e., when
$|A|^{2}<<1$, \ equation(56) reduces to

\begin{equation}\label{e63}
i\frac{\partial A}{\partial \xi}+\frac{1}{2}\frac{\partial^2
A}{\partial
s^2}-\frac{\beta(1+\rho)}{(1+\sigma+\rho)}(1+\sigma-\sigma |A|^2)A
=0.
\end{equation}

The dark soliton solution of above equation turns out to be

\begin{equation}\label{e64}
A(s,\xi)=\rho^{1/2}\tanh\left[-\left(\frac{\beta\rho\sigma}{1+\sigma+\rho}\right)^{1/2}s\right]exp\left[\frac{i\beta(1+\rho)(\rho\sigma-\sigma-1)\xi}{1+\sigma+\rho}\right].
\end{equation}

The spatial width ( $\tau_{FWHM}$) of these solitons are
$\tau_{FWHM}=2\ln(1+\sqrt{2}) \left(\frac{1+ \rho+
\sigma}{-\rho\sigma\beta}\right )^{1/2}$. Rare earth doped strontium
barium niobate (SBN) could be a good candidate for observing these
solitons, since, they have an intermediate level for the two step
excitation. In addition to bright and dark solitons, equation(56)
also predict steady state gray solitons under appropriate bias
condition. These screening gray solitons were investigated by Zhang
et. al.~\citep{96}. Properties of these gray solitons  are similar
to fundamental properties of one-photon photorefractive gray spatial
solitons. For example, they require bias field in opposite to the
optical c-axis of the medium and their FWHM is inversely
proportional to the square root of the absolute value of the bias
field. The main difference between one-photon and two-photon gray
solitons is that one-photon gray solitons rely on one-photon
photorefractive phenomenon to set up space charge field, while the
two-photon photorefractive gray solitons rely on two-photon
photorefractive phenomenon to set up space charge field.

\subsection{Photovoltaic Solitons }

 We consider a photovoltaic
photorefractive crystal under open circuit condition ($R=0$, $g=0$),
 the expression for  space charge  field from equation(53)
reduces to

\begin{equation}\label{e65}
E_{sc}=-E_{p}\frac{s_{2}I_{2}(I_{2}+I_{2d}+\frac{\gamma_{1}N_{A}}{s_{2}})}{(I_{2
}+I_{2d})(I_{1}s_{1}+\beta_{1})}-\frac{D}{\mu
s_{2}}\frac{\gamma_{1}N_{A}}{(I_{2
}+I_{2d})\left(I_{2\infty}+I_{2d}+\frac{\gamma_{1}N_{A}}{s_{2}}\right)}\frac{\partial
I_{2}}{\partial x}.
\end{equation}\

Upon neglecting diffusion,   the nonlinear Schr\"{o}dinger equation
for the normalized envelope    $A(s,\xi)$ can be obtained as

\begin{equation}\label{e66}
i\frac{\partial A}{\partial\xi }+\frac{1}{2}\frac{\partial^{2}
A}{\partial
s^{2}}+\alpha\eta\frac{(1+\sigma+|A|^{2})|A|^{2}}{(1+|A|^{2})}A =0.
\end{equation}

Bright, dark and gray photovoltaic solitons of equation(66) have
been examined by several authors~\citep{93,100}. Deflection of these
solitons and higher order effects  have been also given adequate
attention~\citep{93,100}. For bright solitons, we consider a similar
profile as it was considered for screening solitons, and, the
profile of such solitons turns out to be~\citep{93,100}

\begin{equation}\label{e67}
s=\pm(\alpha\eta)^{-1/2}\int_{y}^{1}\left\{\frac{2\sigma}{p}[\ln(1+p\widehat{y}^{2})-\widehat{y}^{2}\ln(1+p)]+p\widehat{y}^{2}(1-p\widehat{y}^{2})\right\}^{-1/2}d\widehat{y},
\end{equation}

where the nonlinear phase shift is given by
\begin{equation}\label{e68}
\nu=\alpha\eta\sigma[1-\frac{1}{p}\ln(1+p)]+\frac{\alpha\eta p}{2}.
\end{equation}

The bright soliton requires     $\alpha>0, i.e., E_{p}>0$, thus the
photovoltaic field should be in the same direction with respect to
the optical c-axis of the medium. In the low amplitude limit
($|A|^{2}<<1$),  FWHM of these solitons are inversely proportional
to the
square root of the absolute value of the photovoltaic field.\\\\\\
The profile of dark photovoltaic solitons, following a similar
procedure, turns out to be~\citep{93}
\begin{equation}\label{e69}
s=\pm(-\alpha\eta)^{-1/2}\int_{y}^{0}\left[\frac{2\sigma}{1+\rho}(\widehat{y^{2}}-1)-\frac{2\sigma}{\rho}\ln\left(\frac{1+\rho\widehat{y}^{2}}{1+\rho}\right)+\rho(\widehat{y}^{2}-1)^{2}\right]^{-1/2}d\widehat{y},
\end{equation}\
and the nonlinear phase shift is given by

\begin{equation}\label{e70}
\mu=\alpha\eta\rho\left(1+\frac{\sigma}{1+\rho}\right).
\end{equation}\\\
 \  The two-photon photovoltaic solitons require a separate gating
beam to produce a quantity of excited electrons from the valance
band to the intermediate level of the material. Without the gating
beam, the signal beam cannot evolve into a spatial soliton. By
adjusting the gating beam, one can control the width as well as
formation of two-photon photovoltaic solitons.\

\subsection{Screening Photovoltaic Solitons}\
 Steady state screening photovoltaic solitons are obtainable
when an electric field is applied to a photovoltaic photorefractive
crystal. These SP solitons result from both the photovoltaic effect
and spatially nonuniform screening of the applied field. Recently,
bright and dark screening photovoltaic (SP) solitons have been
investigated by Zhang and Liu~\citep{94}. The normalized bright
field profile $y(s)$ of these solitons can be determined ( with \
$\delta=0 $) from equation(54), which turns out to be~\citep{94}:

\begin{eqnarray}\label{e71}
s&=&\pm\int_{y}^{1}\{\frac{2g\beta\sigma}{p(1+\sigma)}[\ln(1+p\widehat{y}^{2})-\widehat{y}^{2}\ln(1+p)]\nonumber \\
& & {
}+\frac{2\alpha\eta\sigma}{p}[\ln(1+p\widehat{y}^{2})-\widehat{y}^{2}\ln(1+p)]
+\alpha\eta
p\widehat{y}^{2}(1-\widehat{y}^{2})\}^{-1/2}d\widehat{y}.
\end{eqnarray}

 The nonlinear phase shift   $\nu$ of these solitons is given by

\begin{equation}\label{e72}
\nu=-\frac{g\beta}{1+\sigma}[1+\frac{\sigma}{p}\ln(1+p)]+\alpha\eta\sigma[1-\frac{1}{p}\ln(1+p)]+\frac{\alpha\eta
p}{2}.
\end{equation}

Please note that unlike bright screening solitons, in the present
case, it is not necessary that the value of $\beta$ should be
positive. However, the sign of  $\alpha$  and $\beta$ should be such
that the curly bracketed term in equation(71) is positive. From
equation(54) the normalized dark field profile y(s) can be obtained
as~\citep{94}:

\begin{eqnarray}\label{e73}
s&=&\pm\int_{y}^{0}\{-\frac{2g\beta\sigma}{(1+\rho+\sigma)}[(\widehat{y}^{2}-1)\nonumber \\
& & {
}-\frac{1+\rho}{\rho}\ln\left(\frac{1+\rho\widehat{y}^{2}}{(1+\rho)}\right)]
-\alpha\eta[\frac{2\sigma(1+g\rho)}{1+\rho}(\widehat{y}^{2}-1)\nonumber \\
& & {
}-\frac{2\sigma(1+g\rho)}{\rho}\ln\left(\frac{1+\rho\widehat{y}^{2}}{1+\rho}\right)+\rho(\widehat{y}^{2}-1)^{2}]\}^{-1/2}d\widehat{y},
\end{eqnarray}

where the nonlinear phase shift is given by

\begin{equation}\label{e74}
\mu=-g\beta+\alpha\eta
\rho(1-g)\left(1+\frac{\sigma}{1+\rho}\right).
\end{equation}
By setting $\alpha=0$ and    $g=1$ in equation(71), we recover
bright solitons of equation(58). Similarly, by setting $\alpha=0$
and $g=1$ in equation(73), we recover dark solitons of equation(62).
In addition, by taking $\beta=0$ and $g=0$ in equation(71) i.e., in
open circuit realization, we recover bright PV solitons of
equation(67). Similarly, by setting $g=0$ and $\beta=0$, we recover
dark PV solitons of equation(69) from equation(73).\

As pointed out by Zhang and Liu~\citep{94}, these   two-photon SP
solitons may be considered as the unity form of two-photon screening
and two-photon photovoltaic solitons under open circuit realization.
If the biased field is much stronger in comparison to the
photovoltaic field, then the screening photovoltaic solitons are
just like screening solitons. If the applied field is absent, the
screening photovoltaic solitons degenerate into the photovoltaic
solitons in the open circuit condition$(\beta=0,  g=0)$. In other
words, the open circuit photovoltaic solitons or screening solitons
are special cases of the screening photovoltaic solitons.
Equation(73) also predicts the existence of two-photon photovoltaic
solitons when $g=1$ and $\beta=0$ i.e.,  two-photon photovoltaic
solitons in closed circuit realization.\\\ Before closing this
section, a brief comment on gray two-photon screening PV  solitons
in biased two-photon phototvoltaic crystals seems inevitable.
Equation(54) predicts the existence of such solitons~\citep{102}.
The properties of these gray solitons, such as, their normalized
intensity profiles, intensity FWHM, transverse velocity and
transverse phase profiles have been discussed in detail by Zhang et.
al.~\citep{102}. They become narrower as the grayness parameter m
decreases for a given normalized  intensity ratio $\rho$. However,
the soliton width generally decreases and transverse velocity
generally increases with intensity ratio $\rho$. In addition,
soliton phase varies in a very involved fashion across transverse
direction and the total phase jump of these solitons exceeds $\pi$
for relatively low value of the grayness parameter $m$.\

\section{ Vector Solitons }
Thus far, we have discussed optical spatial solitons which are
solutions of a single NLS equation. These solutions are  due to a
single optical beam with a  specific polarization and the
polarization is maintained during propagation. However, always this
specific picture  may not hold good. Two or more optical beams may
be mutually trapped and depend on each other in such a way that each
of them propagates undistorted. Thus, several field components at
different or same frequencies or polarizations may interact and
yield shape preserving propagation. In order to discuss such cases,
we need to solve a set of coupled NLS equations. Shape preserving
solutions of this set of coupled NLS equations are called vector
solitons. Only in specific cases, the constituents of these solitons
are vector fields associated with solitons. In general, they are
multi component in nature.

\subsection{Two Component  Incoherently Coupled  Vector Solitons}
\ Among spatial solitons interaction, pairing of two spatial
solitons has been always an intriguing and extensively investigated
issue. When two such soliton forming beams propagate, they interact
through cross phase modulation (XPM) and induce a refractive-index
modulation created by both beams.  Two beams are mutually trapped
and depend on each other in such a way that each of them propagates
undistorted.  Very recently, vector screening
solitons~\citep{104,105,106,107}  have been investigated, that
involve  two polarization components of an optical beam which are
orthogonal to each other. Depending on the symmetry class of the
crystal and its orientation, these solitary beams obey cross or self
coupled vector systems of dynamical equations. \\\ A new type of
steady state incoherently coupled soliton pair was discovered in
biased photorefractive crystals~\citep{108}, which exists only when
the two soliton forming beams possess same polarization and
frequency and are mutually
incoherent~\citep{108,109,110,111,112,113,114,115,116,117,118,119,120,121,122,123}
. These solitons can propagate in bright-bright, dark-dark,
bright-dark and gray-gray configurations, and they can be realized
in simple experimental arrangement with two mutually incoherent
collinearly propagating optical beams. Since two beams are mutually
incoherent, no phase matching is required and they experience equal
effective electro-optic coefficients. The idea of two incoherently
coupled solitons has been generalized and extended to soliton
families where number of constituent solitons are more than two.
Such incoherently coupled families can be established provided they
have same polarization, wavelength and are mutually incoherent.
Bright-bright, dark-dark~\citep{118},
bright-dark~\citep{119,120,121} as well as gray-gray~\citep{113}
configurations have been investigated. These multi component
solitons are stable. In next few  sections, we confine our
discussion on two component spatial photorefractive vector solitons
which are co-propagating and overlapping.\
\subsubsection{Coupled Solitary Wave Equations Due to Single Photon
Phenomenon}
  To start with, we consider a pair of optical beams which are
propagating in a photorefractive crystal ( the crystal could be
PV-PR or non PV-PR) along z-direction. They are of same frequency
and  mutually incoherent. The optical c-axis of the crystal is
oriented along the x direction. The polarization of both beams is
assumed to be parallel to the x-axis. These two optical beams are
allowed to diffract only along the x-direction and y-dynamics has
been implicitly omitted in the analysis. For the sake of simplicity
the photorefractive material is assumed to be lossless. The
perturbed refractive index along the x-axis is given by
$\widehat{n}_{e}^{2}=n_{e}^{2}-n_{e}^{4} r_{33} E_{sc}$.   The
optical fields are expressed in the form $\overrightarrow{E_{1}}
=\overrightarrow{x}\Phi_{1}(x,z)\exp(ikz)$ and
$\overrightarrow{E_{2}} =\overrightarrow{x}\Phi_{2}(x,z)\exp(ikz)$,
where $\Phi_{1}$ and $\Phi_{2}$ are slowly varying envelopes of two
optical fields, respectively. It can be readily shown that the
slowly varying envelopes of two interacting spatial solitons inside
the photovoltaic PR crystal are governed by the following evolution
equations:

\begin{equation}\label{e75}
i\frac{\partial \Phi_{1}}{\partial
z}+\frac{1}{2k}\frac{\partial^{2}\Phi_{1}}{\partial
x^{2}}-\frac{k_{0}n_{e}^{3}r_{33}E_{sc}}{2}\Phi_{1}=0,
\end{equation}

\begin{equation}\label{e76}
i\frac{\partial \Phi_{2}}{\partial
z}+\frac{1}{2k}\frac{\partial^{2}\Phi_{2}}{\partial
x^{2}}-\frac{k_{0}n_{e}^{3}r_{33}E_{sc}}{2}\Phi_{2}=0.
\end{equation}

For relatively broad optical beams  and under strong bias condition,
the space charge field  can be obtained from  equation(11) as
\begin{equation}\label{e77}
E_{sc}=E_{0}\frac{I_{\infty}+I_{d}}{I+I_{d}}+E_{P}\frac{I_{\infty}-I}{I+I_{d}},
\end{equation}

 where in the present case $I=I(x,z)$ is total power density of two optical beams,  $I_{\infty}$
 is the total power density of soliton pair at a distance far away from the center of the crystal
  i.e., $I_{\infty}=I(x\rightarrow \pm\infty)$.   $E_{0}$ is the value of
   the space charge field at far away from the beam center
    i.e., $x \rightarrow \pm\infty$. For two mutually incoherent beams, total optical power density I can be written as
$I(x,z)=n_{e}/(2\eta_{0} )(|\Phi_{1}|^{2}+|\Phi_{2}|^{2})$   i.e.,
sum of Poynting fluxes. Substituting the expression of  $E_{sc}$ in
equations(75) and (76), we derive the following dimensionless
dynamical equations for two soliton forming optical beams:

\begin{eqnarray}\label{e78}
i\frac{\partial A_{j}}{\partial\xi }+\frac{1}{2}\frac{\partial^{2}
A_{j}}{\partial s^{2}}-\beta(1+\rho)
\frac{A_{j}}{(1+|A_{j}|^{2}+|A_{3-j}|^{2})}-\alpha\frac{(\rho-|A_{j}|^{2}-|A_{3-j}|^{2})A_{j}}{(1+|A_{j}|^{2}+|A_{3-j}|^{2})}
=0, \ \ \ \ \j=1,2\ ,\
\end{eqnarray}

where  $\beta=(k_{0}x_{0})^{2}(n_{e}^{4} r_{33}/2) E_{0}$ ,
$\alpha=(k_{0} x_{0})^{2} (n_{e}^{4} r_{33}/2) E_{p}$,\
$A_{j}=\sqrt{\frac{n_{e}}{2\eta_{0}I_{d}}}\Phi_{j}$;\
 $\xi$, $s$ and
$\rho$ are defined earlier. Above set of two coupled Schr\"{o}dinger
equations can be examined for bright-bright, bright-dark, dark-dark,
gray-gray screening, photovoltaic as well as screening photovoltaic
solitons~\citep{110,111,112,113,114,115,116,117,118,119,120}. It can
be easily shown that when the total intensity of the two coupled
solitons is much lower than the effective dark irradiance, the
coupled soliton equations reduce to Manakov equations. The
dark-dark, bright-bright and dark-bright soliton pair solutions of
these Manakov equations can be obtained under appropriate bias  and
photovoltaic fields~\citep{117}.
\\\ With the growing applications of self focusing and spatial solitons in modern
technology, several mathematical methods have evolved to address
soliton dynamics~\citep{124,125,126,127,128,129}. In particular,
Christodoulides et. al.~\citep{108} have developed a very efficient
method to numerically solve a set of coupled equations which has
been employed extensively to investigate coupled solitons in PR
media. In this method, two coupled equations are converted to one
ordinary differential equation which is then numerically solved to
obtain soliton profiles. The main difficulty with this method is its
inability to capture the existence of a large family of stable
stationary solitons. We will discuss more on this in the latter part
of the article.  What follows in  the next section is a discussion
on incoherently coupled solitons employing the method of reference 108.\\\

\subsection{ Incoherently Coupled Screening  Vector Solitons}

 In this section
we consider incoherently coupled bright-bright, dark-dark as well as
bright-dark screening solitons in a biased nonphotovoltaic
photorefractive crystal. These coupled solitons were identified by
Christodoulides et. al.~\citep{108}. The parameter $\alpha=0$,
since, the crystal is nonphotovoltaic, hence, relevant coupled
Schr\"{o}dinger equations  are as follows

\begin{eqnarray}\label{e79}
i\frac{\partial A_{j}}{\partial\xi }+\frac{1}{2}\frac{\partial^{2}
A_{j}}{\partial s^{2}}-\beta(1+\rho)
\frac{A_{j}}{(1+|A_{j}|^{2}+|A_{3-j}|^{2})} =0,
\end{eqnarray}

We first consider a bright-bright soliton pair for which
$I_{\infty}=\rho=0$. Expressing stationary soliton solutions of the
form $A_{1}=p^{1/2} y(s)\cos\theta\ \exp(i\mu\xi)$   and
$A_{2}=p^{1/2} y(s)\sin\theta\ \exp(i\mu\xi)$, where $\mu$
represents nonlinear shift of the propagation constant, and  y(s) is
a normalized real function between $0\leq y(s)\leq 1$. The parameter
$\theta$ is an arbitrary projection angle which ultimately decides
the relative power of two components. Substituting  $A_{1}$ and
$A_{2}$ in equation(79), we get the following ordinary differential
equation

\begin{eqnarray}\label{e80}
\frac{d^{2}y}{ds^{2}}-2\mu y-2\beta\frac{y}{1+py^{2}}=0,
\end{eqnarray}
and
\begin{eqnarray}\label{e81}
\mu=-(\frac{\beta}{p})\ln(1+p).
\end{eqnarray}

Earlier in section (7.3),  it was shown that, above equation admits
bright solitons when $\beta$ \ i.e., $E_{0}$    is positive. The,
same condition holds good for bright-bright pair which can be
obtained by numerically solving equation (80). A typical
bright-bright pair has been depicted in figure (9). \

\begin{center}
\textit{Insert Figure (9) here}
\end{center}

For dark-dark pairs, $I_{\infty}$ and $\rho$ are finite. We express
$A_{1}$  and $A_{2}$ as $A_{1}=\rho^{1/2} y(s)\cos\theta
\exp(i\mu\xi)$   and $A_{2}=\rho^{1/2} y(s)\sin\theta
\exp(i\mu\xi)$, with $|y(s)|\leq 1$. Thus, we have

\begin{eqnarray}\label{e82}
\frac{d^{2}y}{ds^{2}}-2\mu y-2\beta(1+\rho)\frac{y}{1+\rho y^{2}}=0,
\end{eqnarray}
and
\begin{eqnarray}\label{e83}
\mu=-\beta.
\end{eqnarray}
Equation (82) can be solved for dark-dark soliton pairs provided
$\beta $ \ i.e., $E_{0}$    is negative. A typical dark-dark pair
has been depicted in figure (10).\

\begin{center}
\textit{Insert Figure (10) here}
\end{center}

 For bright-dark soliton pair, we express $A_{1}$  and   $A_{2}$   as $A_{1}=p^{1/2} f(s)\exp(i\mu\xi)$   and
  $A_{2}=\rho^{1/2} g(s)\exp(i\nu\xi)$, where $f(s)$  and $g(s)$  respectively represents envelope of bright and dark beams.
   Two positive quantities  $p$  and $\rho$   represent the ratios of their maximum power density with respect to the dark
   irradiance   $I_{d}$.  Therefore, bright-dark soliton pair obeys following coupled ordinary differential
   equations:

\begin{eqnarray}\label{e84}
\frac{d^{2}f}{ds^{2}}-2\left[\mu +
\frac{\beta(1+\rho)}{1+pf^{2}+\rho g^{2}}\right]f=0,
\end{eqnarray}
and
\begin{eqnarray}\label{e85}
\frac{d^{2}g}{ds^{2}}-2\left[\nu +
\frac{\beta(1+\rho)}{1+pf^{2}+\rho g^{2}}\right]g=0.
\end{eqnarray}

A particular solution of above equations can be obtained using the
simplification   $f^{2}+g^{2}=1$. Employing appropriate boundary
conditions, the nonlinear phase shifts $\mu$  and $\nu$  are
obtained as: $\mu=-\frac{\beta}{\Lambda} \ln(1+\Lambda)$  and
$\nu=-\beta$, where $ \Lambda= (p-\rho)/(1+\rho)$. When peak
intensities of  two solitons are approximately equal, $\Lambda<<1$ .
The approximate soliton solution~\citep{105,108}  in this particular
case is given by\ $A_{1}=p^{1/2}\
sech[(\beta\Lambda)^{1/2}s]\exp[-i\beta(1-\Lambda/2)\xi],$ and
$A_{2}=\rho^{1/2}\tanh[(\beta\Lambda)^{1/2}s]\exp[-i\beta\xi].$ \
These two soliton solutions are possible only when the product
$(\beta\Lambda)$ is a positive quantity. A typical bright dark
soliton pair is depicted in figure (11).

\begin{center}
\textit{Insert Figure (11) here}
\end{center}

\subsubsection{ Missing Bright Screening PV Solitons}

In the preceeding section, we have followed the procedure developed
in reference 108 and assumed a particular type of ansatz for the
bright-bright pair. According to this method, the normalized power
$P_{1}$ and $P_{2}$ of two solitons of the bright-bright pair are
$P_{1}=p\sin^{2}\theta$ and $P_{2}=p\cos^{2}\theta$.  Therefore, for
a given $P_{1}$, $\theta$ has a fixed value, hence, the power of the
other component has only one possible value which is unique. Or in
other words a composite soliton can exist only with a single power
ratio. In this section, we will show that for a given power of one
component, the other component can exist with different power. Thus,
the method of reference~\citep{108}  fails to identify a large
number of bright-bright solitons in a two component configuration.\
Therefore, our goal, in this section,  is to demonstrate the
existence of a new very large family of two-component composite
screening photovoltaic spatial solitons
  in biased photovoltaic-photorefractive crystals which were not identified by the method of reference 108. For bright solitons, $\rho=0$,
  hence, relevant coupled equations for bright-bright screening PV solitons in biased PV-PR crystals~\citep{64}
  are

\begin{eqnarray}\label{e86}
i\frac{\partial A_{j}}{\partial\xi }+\frac{1}{2}\frac{\partial^{2}
A_{j}}{\partial
s^{2}}+\frac{[\alpha(|A_{j}|^{2}+|A_{3-j}|^{2})-\beta]A_{j}}{(1+|A_{j}|^{2}+|A_{3-j}|^{2})}
=0,
\end{eqnarray}

In order to analyze the behavior of these coupled solitons, we
assume  solutions of the form

\begin{equation}\label{e87}
A_{j}(s,\xi)=\Psi_{j}(s,\xi)\exp(-i\Omega_{j}\xi),
\end{equation}

By virtue of use of equation(87) in equation (86), we obtain
following equations:

\begin{equation}\label{e88}
\frac{\partial
\Omega_{j}}{\partial\xi}\Psi_{j}-\frac{1}{2}\left(\frac{\partial
\Omega_{j}}{\partial
s}\right)^{2}\Psi_{j}+\frac{1}{2}\frac{\partial^{2}\Psi_{j}}{\partial
s^{2}}+\alpha\Pi\Psi_{j}=0,
\end{equation}

\begin{equation}\label{e89}
\frac{\partial \Psi_{j}}{\partial\xi}-\frac{\partial
\Omega_{j}}{\partial s}\frac{\partial\Psi_{j}}{\partial
s}-\frac{1}{2}\frac{\partial^{2}\Omega_{j}}{\partial
s^{2}}\Psi_{j}=0,
\end{equation}

where \
$\Pi=\frac{\alpha(|\Psi_{1}|^{2}+|\Psi_{2}|^{2})-\beta}{1+|\Psi_{1}|^{2}+|\Psi_{2}|^{2}}$.
\ The last three terms of (88)  determine the behavior of the
eikonal $\Omega_{j}$    i.e., the convergence or divergence of two
optical beams. The fourth term in this  equation represents
nonlinear refraction while the third term determines diffraction.
Equation (89) determines the evolution of the beam envelope
$\Psi_{j}$.  In equation (88), $\Pi$ represents the contribution
from nonuniform screening of the applied electric field and
photovoltaic properties of the crystal as well.
\\\\ Lowest order localized bright solitons, for which light is
confined in the central region of the soliton, obey $\Psi_{j}
(s=0)=\Psi_{jmax}$  and $\Psi_{j}=0$ as $ s\rightarrow \pm \infty.$
The fundamental solutions of coupled Schr\"{o}dinger equations, in a
self focusing Kerr  medium, are represented by sech functions.
Equation (86) is a modified nonlinear Schr\"{o}dinger
equation(MNLSE) in saturating media. Due to the saturating nature of
the medium, it is expected that the fundamental soliton solutions
will not be exactly sech function. However, in many nonlinear
optical problems involving NLSE and MNLSE in Kerr, cubic quintic,
nonlocal and saturating media, approximate solutions have been
obtained using Gaussian ~\citep{130,131,132,133,134,135} or super
Gaussian ansatz~\citep{135}. The motivation of employing such ansatz
is two fold. Firstly, particularly true for Gaussian ansatz,
mathematical formulations become easy. Secondly, numerically
computed exact solutions are not widely different from Gaussian
profiles in many cases. Thus, though approximate, Gaussian profile
still provides good approximation to the problem. Hence, the
solutions of above equations are  taken to be Gaussian with
amplitude and phase of the following form

\begin{equation}\label{e90}
\Psi_{j}(s,\xi)=\frac{\Psi_{j0}}{\sqrt{f_{j}(\xi)}}exp\left[-\frac{s^{2}}{2r_{j}^{2}(\xi)f_{j}^{2}(\xi)}\right],
\end{equation}

\begin{equation}\label{e91}
\Omega_{j}(s,\xi)=\frac{s^{2}}{2}\beta_{j}(\xi)+\phi_{j}(\xi),
\end{equation}\
and
\begin{equation}\label{e92}
\beta_{j}(\xi)=-\frac{1}{f_{j}}\frac{df_{j}(\xi)}{d\xi},\ \ \ \
\end{equation}

where,   $\Psi_{jo}$         represents the peak power of the
component of bright-bright solitons, $r_{j}$ is a positive constant,
$f_{j} (\xi)$  is  variable spatial width parameter; $r_{j} f_{j}
(\xi)$ is  the spatial width of these solitons  and  $\phi_{j}
(\xi)$ is an arbitrary phase function. For a pair of nondiverging
solitons at $\xi=0$, we should have $f_{j}=1$ and
$\frac{df_{j}}{d\xi}=0$. Substituting for $\Psi_{j}$  and
$\Omega_{j}$ in (88), using paraxial ray
approximation~\citep{86,126,127}  and equating coefficients of
$s^{2}$ from  both sides of  (88), we obtain

\begin{equation}\label{e93}
\frac{d^{2}f_{j}}{d\xi^{2}}=\frac{1}{r_{j}^{4}f_{j}^{3}}-C\left(\frac{P_{j}}{r_{j}^{2}f_{j}^{2}}
+\frac{P_{3-j}}{r_{3-j}^{2}f_{3-j}^{3}}\right)\frac{1}{\left(1+\frac{P_{j}}{f_{j}}+\frac{P_{3-j}}{f_{3-j}}\right)^{2}},
\end{equation}\

where  $P_{j}=\Psi_{j0}^{2}$      and $C=2(\alpha+\beta)$. The
solutions of above equation should give stationary and
non-stationary coupled solitons of  (86)  for given set of power and
spatial width.

\subsubsection{Stationary Composite Solitons }
In order to identify stationary composite solitons, we need to
locate equilibrium points. The equilibrium points of equation(93)
 can be obtained from following equations
\begin{equation}\label{e94}
\frac{1}{r_{1}^{4}}-C\left(\frac{P_{1}}{r_{1}^{2}}
+\frac{P_{2}}{r_{2}^{2}}\right)\frac{1}{\left(1+P_{1}+P_{2}\right)^{2}}=0,
\end{equation}
and
\begin{equation}\label{e95}
\frac{1}{r_{2}^{4}}-C\left(\frac{P_{1}}{r_{1}^{2}}
+\frac{P_{2}}{r_{2}^{2}}\right)\frac{1}{\left(1+P_{1}+P_{2}\right)^{2}}=0.
\end{equation}\

From above equations, it is obvious that $r_{1}=r_{2}=r$ is the
condition  for existence of stationary coupled solitons, where $r$
is a constant. Therefore, composite solitons with different spatial
widths cannot propagate as a stationary entity. The existence
equation of coupled solitons turns out to be

\begin{equation}\label{e96}
Cr^{2}=\frac{(1+P_{1}+P_{2})^{2}}{P_{1}+P_{2}}.
\end{equation}\

Obviously, for a bright-bright pair $C$  should be positive. From
the existence equation, it is evident that though spatial widths of
each of the two components are equal, their respective peak power
can have different value. Equation (96)  is equivalent to a
quadratic equation in $P_{1}$, the root of  which  is obtained as

\begin{equation}\label{e97}
P_{1}=\left[(Cr^{2}-2P_{2}-2)\pm\sqrt{(Cr^{2}-2P_{2}-2)^{2}-4(P_{2}^{2}+2P_{2}+1-Cr^{2}P_{2})}\right]/2.
\end{equation}\

$P_{1}$   and $P_{2}$ are  real and positive, hence, always
$Cr^{2}\geq 4$ . For a given value of  $C$, this relationship
dictates a minimum width for the propagating soliton pair.  The
variation of $P_{1}$ with  $P_{2}$ for different values of  $r$ has
been depicted in figure 12(a)-(b). Each point on any curve of these
figures represents a stationary composite soliton with a definite
spatial width and peak power. \\

\begin{center}
\textit{Insert Figure (12) here}
\end{center}

An important issue is, whether for a given peak power of one of the
component, the other component exists with only one or multiple
values of peak power. This issue can be settled from figure (13)
which shows the variation of width with peak power of one component
keeping peak power of other component constant. It is evident from
the figure that $P_{1}$  has a range of values for a fixed $P_{2}$,
thus, with fixed $P_{2}$ other component can exist with different
values of $P_{1}$. At this stage it is worth pointing out that these
solitons cannot be identified with the method employed in ref. 108.

\begin{center}
\textit{Insert Figure (13) here}
\end{center}

\subsubsection{Degenerate Bright Screening PV Bistable Solitons}

 We take up a degenerate case in which peak power of two components is same and
having same spatial width. Setting   $P_{1}=P_{2}=P$ in (97), we
obtain a quadratic equation of  P, the solution of which is obtained
as $P=[Cr^{2}-2\pm\sqrt{(Cr^{2}-4)^{2}-4]}/4$ , which implies that
spatial width $r$ of each component of the composite soliton should
be greater than $2/\sqrt{C}$, i.e., a two-component composite
soliton whose individual spatial width is less than above value
cannot propagate as a self trapped mode. In figure (14) we have
displayed variation of $r$ with peak power $P$. From figure,
existence of a bistable regime~\citep{66}  is evident i.e., two sets
of soliton pairs exist with same spatial width but having different
peak power and consequently different peak amplitude. Only this
degenerate case possesses bistable property.

\begin{center}
\textit{Insert Figure (14) here}
\end{center}
\subsubsection{ Numerical Simulation}

To verify the predictions of foregoing analysis, it is essential to
perform numerical simulation. Equation (86)  has been solved
numerically using the split step Fourier beam propagation method
~\citep{3}. To begin with, we  look for behavior of the soliton at
low power. From figure 12(b), we choose $Cr^{2}=4$ and $r=1/5$ and
select different points from the curve leveled with this value.
Chosen points have following values of $P_{1}$ and $P_{2}$,
particularly, (i)$P_{1}=0.5, P_{2}=0.5$, (ii)$P_{1}=0.6666,
P_{2}=0.3333$, (iii)$P_{1}=0.8333, P_{2}=0.1666$ and
(iv)$P_{1}=0.909, P_{2}=0.0909$. It must be emphasized that each
point corresponds to a stationary composite soliton, a paraxial
theory prediction. With these parameter values, we launch two
Gaussian optical beams
$A_{1}=\sqrt{P_{1}}\exp\left(-\frac{s^{2}}{2r^{2}}\right)$ and
$A_{2}=\sqrt{P_{2}}\exp\left(-\frac{-s^{2}}{2r^{2}}\right)$ in (86).
 Both Gaussian beams acquire solitonic shape asymptotically
without major modification within very small distance and then they
propagate almost as a stationary composite soliton.The behavior of
two Gaussian spatial solitons corresponding
to each of these points has been depicted in figure (15).\\\
\begin{center}
\textit{Insert Figure (15) here}
\end{center}

 It is evident from these figures that a soliton with large power
can trap another soliton whose power is much lower and both can
propagate as a stationary bound state. At this stage it would be
appropriate to cite one practical example. Consider  a $BaTiO_{3}$
crystal  at a wavelength $\lambda_{0}=0.5\mu m$ with following
crystal parameters~\citep{64}: $n_{e}=2.365$, $r_{33}=80\times
10^{-12}m/V,$ $E_{p}=10^{5}V/m$. We take arbitrary spatial scale
$x_{0}=40\mu m$ and  $E_{o}=5.8\times 10^{4}$ V/m. With these
 values, we find, $\beta=18.4,\ \alpha=31.58$ \ and  $C=2(\alpha+\beta)=99.96$. For $Cr^{2}=4, r\approx
0.20$, thus, in natural unit the intensity FWHM ( i.e.,$1.665r$) of
two components is found to be $13.32\mu m$ . An important point to
note is that the present investigation is also valid for
$LiNbO_{3}$. The $LiNbO_{3}$ parameters could be taken as
$n_{e}=2.2, r_{33}=30\times 10^{-12}$m/V and  $|E_{p}|=4\times
10^{6}$ V/m. However, it should be pointed out that, while $E_{p}$
could be either positive or negative for $BaTiO_{3}$, depending on
the polarization of light, the experimental results show that
$E_{p}$ is always negative for $LiNbO_{3}$. Therefore, with proper
choice of the value of  $E_{o},\ C=2(\alpha+\beta)$ can be made
positive, and hence bright-bright
coupled soliton pairs is also  observable in $LiNbO_{3}$.\\

We now proceed to obtain bistable composite solitons numerically.
From figure  (14) we choose two points A and B, each corresponds to
one pair of composite soliton. These two points have equal values of
$Cr^{2}(=4.2)$   but two different values of soliton peak power $P$.
We have taken $C=50$, hence, $r=0.290$  for both A and B.  Peak
power P for points A and B are  $0.321$  and $0.779$, respectively.
Numerically obtained dynamic evolution of a pair of composite
bistable solitons corresponding to above values has been
demonstrated in figure (16).\
\begin{center}
\textit{Insert Figure (16) here}
\end{center}\
 \\In the figure, the upper panel
represents one composite soliton while the lower panel represents
another. From figure, it is evident that both components of the
composite in the upper panel remain absolutely stationary as they
propagate, an example where paraxial theory prediction complies with
high accuracy. However, in the lower panel, both components of the
composite propagate as stable self trapped mode though they keep on
gentle breathing. Hence, in this case, though prediction of paraxial
approximation is not very accurate, yet it is able to capture the
overall features broadly. Finally,  we have found that bright-bright
pairs are stable against small perturbation in peak amplitude and
spatial width. Both paraxial theory and numerical simulation show
that identified composite solitons are stable.\\\

\subsection{Incoherently Coupled  Solitons Due to  Two-Photon
Photorefractive Phenomenon}

 To investigate coupled solitons in two-photon
photorefractive media, the required optical configuration is very
similar to the one discussed in sec 8.1. The only difference between
this case and the earlier one is that, in the present case there are
two incoherent soliton forming beams whose polarization and
frequencies are same, whereas in the former case there is only one
soliton forming beam. As usual,   the optical fields are expressed
in the form
$\overrightarrow{E_{1}}=\overrightarrow{x}\Phi_{1}(x,z)\exp(ikz)$
and
$\overrightarrow{E_{2}}=\overrightarrow{x}\Phi_{2}(x,z)exp(ikz)$,
where  $\Phi_{1}$  and $\Phi_{2}$  are slowly varying envelopes of
two optical fields, respectively.  The coupled Schr\"{o}dinger
equations for the normalized slowly varying envelopes of two optical
fields can be described as

\begin{equation}\label{e98}
i\frac{\partial U}{\partial\xi }+\frac{1}{2}\frac{\partial^{2}
U}{\partial s^{2}}-g\beta
\frac{(1+\rho)(1+\sigma+|U|^{2}+|V|^{2})U}{(1+\sigma+\rho)(1+|U|^{2}+|V|^{2})}-\alpha\frac{\eta(g\rho-|U|^{2}-|V|^{2})(1+\sigma+|U|^{2}+|V|^{2})U}{(1+|U|^{2}+|V|^{2})}
=0,
\end{equation}
and

\begin{equation}\label{e99}
i\frac{\partial V}{\partial\xi }+\frac{1}{2}\frac{\partial^{2}
V}{\partial s^{2}}-g\beta
\frac{(1+\rho)(1+\sigma+|U|^{2}+|V|^{2})V}{(1+\sigma+\rho)(1+|U|^{2}+|V|^{2})}-\alpha\frac{\eta(g\rho-|U|^{2}-|V|^{2})(1+\sigma+|U|^{2}+|V|^{2})V}{(1+|U|^{2}+|V|^{2})}
=0,
\end{equation}

where $U=\sqrt{\frac{n_{e}}{2\eta_{0}I_{2d}}}\Phi_{1}$,
$V=\sqrt{\frac{n_{e}}{2\eta_{0}I_{2d}}}\Phi_{2}$; parameters $\rho,\
\xi,\ s,\ \beta,\ \alpha,\ \eta,\ \sigma$ have been defined earlier.\\\\
Incoherently coupled solitons in two-photon photorefractive media
have recently received tremendous attention, since, the dynamics of
these solitons can be controlled by a separate gating
beam~\citep{137,138,139,140,141,142,143,144,145,146,147,148,149,150,151,152}.
Properties of these solitons can be investigated using equations
(98) and (99).In next few sections we will present a brief
description of these solitons.

\subsubsection{Incoherently Coupled Two-Photon Photovoltaic solitons Under
Open Circuit Condition}

 In this section, we discuss the existence and
nonlinear dynamics of two-component incoherently coupled composite
solitons in two-photon photorefractive materials under open circuit
condition. In the steady state regime, these incoherently coupled
solitons can propagate in bright-dark, bright-bright and dark-dark
configurations. These photovoltaic soliton families can be
established provided that the carrier beams share same polarization
and wavelength, and numerical simulations show that these solitons
are stable for small perturbation on amplitude.  For photovoltaic
solitons under open circuit configuration, $g=0$   and $\beta=0$,
hence, relevant Schr\"{o}dinger equations are

\begin{equation}\label{e100}
i\frac{\partial U}{\partial\xi }+\frac{1}{2}\frac{\partial^{2}
U}{\partial
s^{2}}+\eta\alpha\frac{(|U|^{2}+|V|^{2})(1+\sigma+|U|^{2}+|V|^{2})U}{(1+|U|^{2}+|V|^{2})}
=0,
\end{equation}

\begin{equation}\label{e101}
i\frac{\partial V}{\partial\xi }+\frac{1}{2}\frac{\partial^{2}
V}{\partial
s^{2}}+\eta\alpha\frac{(|U|^{2}+|V|^{2})(1+\sigma+|U|^{2}+|V|^{2})V}{(1+|U|^{2}+|V|^{2})}
=0.
\end{equation}\ \
\paragraph{ Bright-dark solitons} \ \

We first discuss the properties of  photovoltaic bright-dark soliton
pairs. To obtain the solution for a bright-dark soliton pair, the
normalized envelopes $U$  and $V$  are expressed as

\begin{equation}\label{e102}
U=p^{1/2}f(s)exp(i\mu\xi),
\end{equation}
\begin{equation}\label{e103}
V=\rho^{1/2}g(s)exp(i\nu\xi).
\end{equation}

In above expressions,   $f(s)$ and $g(s)$ are real functions, which
correspond to the bright and dark profile, respectively. These
 real functions are bounded i.e., $0\leq f(s)\leq 1$ and
$0\leq g(s)\leq1$. Also, $p$  and  $\rho$ respectively represents
the ratio of solitons maximum intensity  to the dark irradiance
$I_{2d}$. Inserting expressions (102) and (103) in equations (100)
and (101), we
obtain\\
\begin{equation}\label{e104}
\frac{d^{2}f}{ds^{2}}-2(\mu-\alpha\eta\sigma)f+2\alpha\eta(pf^{2}+\rho
g^{2})f-\frac{2\alpha\eta\sigma f}{1+pf^{2}+\rho g^{2}}=0,
\end{equation}
\begin{equation}\label{e105}
\frac{d^{2}g}{ds^{2}}-2(\nu-\alpha\eta\sigma)g+2\alpha\eta(pf^{2}+\rho
g^{2})g-\frac{2\alpha\eta\sigma g}{1+pf^{2}+\rho g^{2}}=0.
\end{equation}

We look for a particular solution which satisfies the condition
$f^{2}+g^{2}=1$. Nonlinear propagation constants $\mu$  and $\nu$
can be determined using appropriate boundary conditions. The value
of these turn out to be

\begin{equation}\label{e106}
\mu=\alpha\eta\sigma\left[1-\frac{\log(1+\Delta)}{\Delta(1+\rho)}\right]+\frac{\alpha\eta(p+\rho)}{2},
\end{equation}\
and
\begin{equation}\label{e107}
\nu=\alpha\eta\rho\left[1+\frac{\sigma}{1+\rho}\right],
\end{equation}\

where  $\Delta=\left(\frac{p-\rho}{1+\rho}\right)$.\ At this stage a
comment on the sign of $\alpha$  for the existence of bright-dark
solitons is desirable. In order to do that we integrate
equation(104) to obtain

\begin{equation}\label{e108}
s=\pm\frac{1}{(-\alpha\eta)^{1/2}}\int_{1}^{f}\left[\frac{2\sigma}{\Delta(1+\rho)}(f^{2}\log(1+\Delta)-\log(1+\Delta
f^{2}))+(p-\rho)f^{2}(f^{2}-1)\right]^{1/2}df.
\end{equation}\

The sign of the integrand within the third bracket depends on the
parameters $p,\ \rho$ and  $\sigma$. For a given set of
experimentally relevant values  of the aforesaid parameters, for
example,\ when $p=3.95,\rho=4.0$  and $\sigma=10^{6}$, the integrand
within the third bracket is positive. Thus, for above set of
parameters the existence of dark-bright solitons requires $\alpha<0$
i.e., $E_{p}<0 $. For illustration, we consider a $LiNbO_{3}$
crystal with the following parameters $n_{e}=2.2$ and
$r_{33}=30\times10^{-12}$ $mV^{-1}$at wavelength $\lambda_{0}$.
Other parameters are taken as $E_{p}=-4\times10^{6}$  $Vm^{-1}$,\ $
s_{1}=3\times10^{-4}m^{2} W^{-1} s^{-1}$,\
$\gamma_{1}=3.3\times10^{-17} m^{3} s^{-1},\ N_{A}=10^{22} m^{-3},\
\beta_{1}=0.5s^{-1},\ \beta_{2}=0.5s^{-1}$  and
$s_{2}=3\times10^{-4} m^{2} W^{-1} s^{-1}$.  The gating beam
intensity $I_{1}=10^{6} W/m^{2}$, the scaling parameter
$x_{0}=0.5\mu m$, therefore, $\alpha\approx -22.2$ and  $\eta
=1.67\times10^{-4}$. The value of $\sigma$ can be controlled by
modulating the dark irradiance artificially using incoherent
illumination~\citep{93} and for the present investigation we take
$\sigma=10^{6}$   and   $\rho=4$. A typical bright-dark pair has
been depicted in figure (17).\
\begin{center}
\textit{Insert Figure (17) here}
\end{center}

  In order to examine the influence of the gating beam on these
solitons, we have numerically computed profiles of these solitons at
two different values of the gating beam intensities. This has been
depicted in figure (18). With the change in $I_{1}$ , the width of
each component changes.
\begin{center}
\textit{Insert Figure (18) here}
\end{center}


\paragraph{ Bright-bright solitons}\ \ \\
We now investigate  two component bright-bright solitons. In this
case, intensities of both soliton forming optical beams  vanish at
infinity i.e., as   $s\rightarrow\pm\infty,\ I_{2\infty}=0$. The
soliton solution is now expressed in terms of normalized envelopes
$U$ and $V$ as

\begin{equation}\label{e109}
U=p^{1/2}y(s)cos\theta\exp(i\mu\xi),
\end{equation}
\begin{equation}\label{e110}
V=p^{1/2}y(s)sin\theta\exp(i\mu\xi),
\end{equation}
where  $p$      represents the ratio of the peak intensity to the
dark irradiance $I_{2d}  ,\   \mu$  is the nonlinear shift of the
propagation constant, $y(s)$  is the normalized real function which
is bounded as   $0\leq y(s)\leq1$ ,  $\theta$ is an arbitrary
projection angle which describes relative strength of two components
of the composite. Substitution of expressions (109) and (110) in
either of equations (100) or (101) yields the following differential
equation,

\begin{equation}\label{e111}
\frac{d^{2}y}{ds^{2}}=2(\mu-\alpha\eta\sigma)y-2\alpha\eta
py^{3}+\frac{2\alpha\eta\sigma y}{1+py^{2}}=0.
\end{equation}\
Integrating above equation once, we obtain,
\begin{equation}\label{e112}
\left(\frac{dy}{ds}\right)^{2}=2(\mu-\alpha\eta\sigma)(y^{2}-1)-\alpha\eta
p(y^{4}-1)+\frac{2\alpha\eta
\sigma}{p}\log\left(\frac{1+py^{2}}{1+p}\right).
\end{equation}\

Making use of the boundary conditions   $ y(\pm\infty)=0$ and
$\dot{y}(\pm\infty)=0$, we can easily obtain $\mu$ as

\begin{equation}\label{e114}
\mu=\alpha\eta\sigma\left[1-\frac{\log(1+p)}{p}\right]+\frac{\alpha\eta
p}{2}.
\end{equation}\
Inserting equation (113) in (112) we get,

\begin{equation}\label{e114}
\left(\frac{dy}{ds}\right)^{2}=\alpha\{\frac{2\eta\sigma}{p}[\log(1+py^{2})-y^{2}\log(1+p)]+\eta
py^{2}(1-y^{2})\}.
\end{equation}

From equation (114) we can easily show that the quantity within the
curly bracket in the right hand side  is positive for all the values
of $y^{2} (s)$ i.e., $0\leq y(s)\leq 1$, therefore, we easily
conclude that $\alpha
>0$ for  bright-bright solitons. In order to investigate
bright-bright soliton pair, we take a Cu:KNSBN crystal, whose
parameters at $\lambda_{0}=0.5\mu m$   are taken  as $n_{e}=2.2$, $
r_{33}=200\times10^{-12} mV^{-1}$ and $E_{p}=2.8\times10^{6}
Vm^{-1}$. The scaling parameter $x_{0}=10\mu m$,  $p=10$  and
$\theta=30^{0}$. Other parameters for the bright-bright soliton
configuration are: $\alpha=22.2$, $\eta=1.5\times10^{-4}$ and
$\sigma=10^{4}$. Figure (19) depicts the normalized intensity
profile of the photovoltaic bright-bright soliton pair.
\setcounter{secnumdepth}{5} \setcounter{tocdepth}{5}

\begin{center}
\textit{Insert Figure (19) here}
\end{center}

\paragraph{Dark-dark  solitons}\ \ \ \\\
Properties of dark-dark soliton pairs can be analyzed following
similar procedure as elucidated in previous sections. In the case of
dark type profiles, there is a constant intensity background i.e.,\
$I_{2\infty}\neq 0$, therefore we express normalized envelopes $U$
and $V$as
\begin{equation}\label{e115}
U=\rho^{1/2}y(s)cos\theta\exp(i\mu\xi),
\end{equation}
\begin{equation}\label{e116}
V=\rho^{1/2}y(s)sin\theta\exp(i\mu\xi),
\end{equation}\

where,  $\rho=I_{2\infty}/I_{2d}$  and $0\leq y(s)\leq 1$. As usual
$\mu$ is the nonlinear shift of the propagation constant, and
$\theta$ is the projection angle. Furthermore, inserting equations
(115) and (116) in either of equation (100) or (101), we obtain

\begin{equation}\label{e117}
\frac{d^{2}y}{ds^{2}}=2(\mu-\alpha\eta\sigma)y-2\alpha\eta \rho
y^{3}+\frac{2\alpha\eta\sigma y}{1+\rho y^{2}}.
\end{equation}\

Above equation can be solved easily adopting numerical procedure
after evaluationg the nonlinear propagation constant $\mu$. An
important point to note is that  dark-dark soliton pairs require
$\alpha <0$. For illustration, we take a $LiNbO_{3}$ crystal with
$n_{e}=2.2$ and $ r_{33}=30\times10^{-12} mV^{-1}$   at wavelength
$\lambda_{0}=0.5\mu m$. Other parameters are $E_{p}=-4\times10^{6}
Vm^{-1}$, $ s_{1}=3\times10^{-4} m^{2} W^{-1} s^{-1}$,
$\gamma_{1}=3.3\times10^{-17} m^{3} s^{-1}$, $ N_{A}=10^{22}
m^{-3}$, $\beta_{1}=0.5s^{-1}, \beta_{2}=0.5s^{-1}$ and
$s_{2}=3\times10^{-4} m^{2} W^{-1} s^{-1}$. The gating beam
intensity $I_{1}$is taken to be $10^{6} W/m^{2}$ and the scaling
parameter $x_{0}=0.5\mu m$. Therefore, $\alpha\approx -22.2$,
$\eta=1.67\times10^{-4}$. We take $\sigma=10^{6}$  and $\rho=4$.  A
dark-dark soliton pair is depicted in figure (20).  Numerical
simulation confirms  that these solitons are robust, do not break up
or disintegrate if small perturbation in amplitude is introduced.

\begin{center}
\textit{Insert Figure (20) here}
\end{center}

\section{Conclusion }\
 We have presented a brief review of the recent
developments in the field of optical spatial solitons in
photorefractive media. In relatively short time this topic has
achieved tremendous success in theory as well as in experiments. We
have considered fundamental properties of three types of solitons,
particularly, screening, photovoltaic and screening photovoltaic
solitons and described different methods to investigate them. For
each type of soliton, three different configurations i.e., bright,
dark and gray varieties have been considered. Self bending of these
solitons due to diffusion and effect of higher order diffusion on
self bending phenomenon are also highlighted. Besides single photon
photorefractive phenomenon, lately the two-photon photorefractive
phenomenon has become a topic of intense research since the PR
effect can be controlled with a separate gating beam. Mechanisms of
formation of PR solitons due to single photon as well as two-photon
photorefractive processes have been discussed. Interaction of
solitons is an extremely important topic which could be exploited to
fabricate all optical switching devices. We have discussed important
properties associated with interaction of these solitons. Vector
solitons, particularly, incoherently coupled solitons due to single
photon and two-photon photorefractive phenomena have been
highlighted. Existence of some missing solitons pointed out. Several
properties discovered so far for these solitons are universal and
applicable to other branches of solitons.

\end{onehalfspacing}



\section*{Acknowledgment}

Part of the work was done under the framework of Senior Associate
scheme of the Abdus Salam International Center for Theoretical
Physics(ICTP), Italy. One of the authors, SK, thank ICTP for the
warm hospitality extended under the aforementioned scheme. He would
also like to thank Prof. Ajoy Chackraborty, Vice Chancellor, Birla
Institute of Technology, Meara, Ranchi, for encouragement.

\newpage

\begin{figure}
\centering \scalebox{0.5}
{\includegraphics[height=10.5in,width=10.5in]{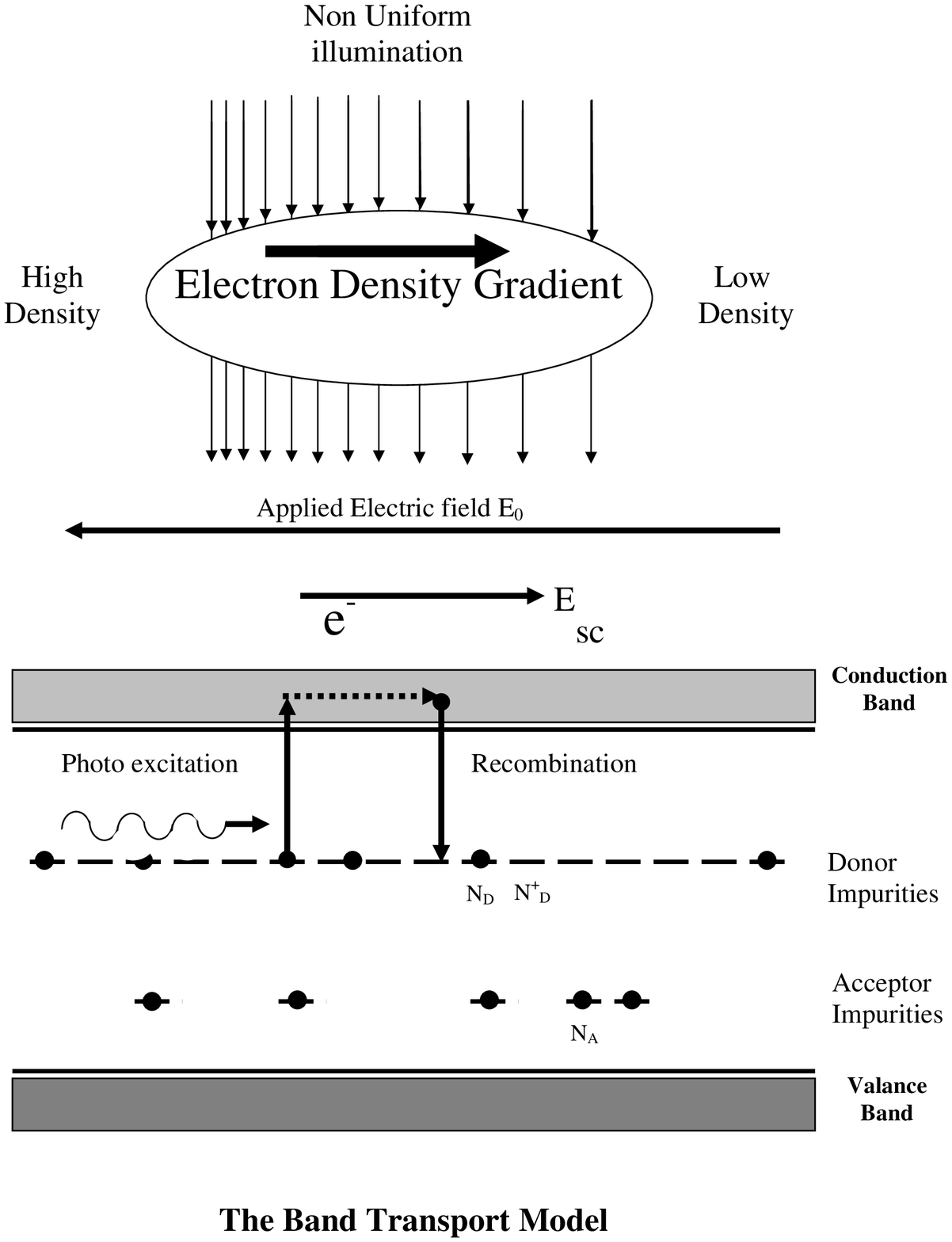}}
\caption{Band transport model} \label{fig:fig1}
\end{figure}
\newpage

\begin{figure}
\centering \scalebox{0.5}
{\includegraphics[height=6.5in,width=6.5in]{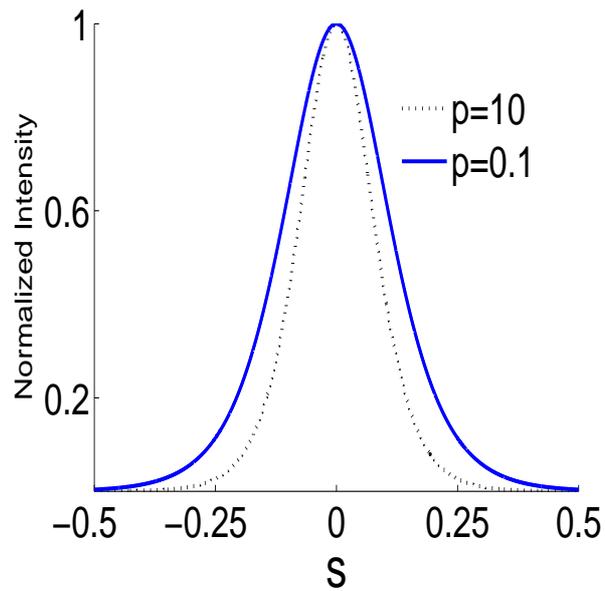}}
\caption{Normalized intensity profile of bright spatial solitons for
$\beta=43, x_{0}=20\mu m; p=0.1$  and $ 10$.} \label{fig:fig2}
\end{figure}

\begin{figure}
\centering \scalebox{0.5}
{\includegraphics[height=6.5in,width=6.5in]{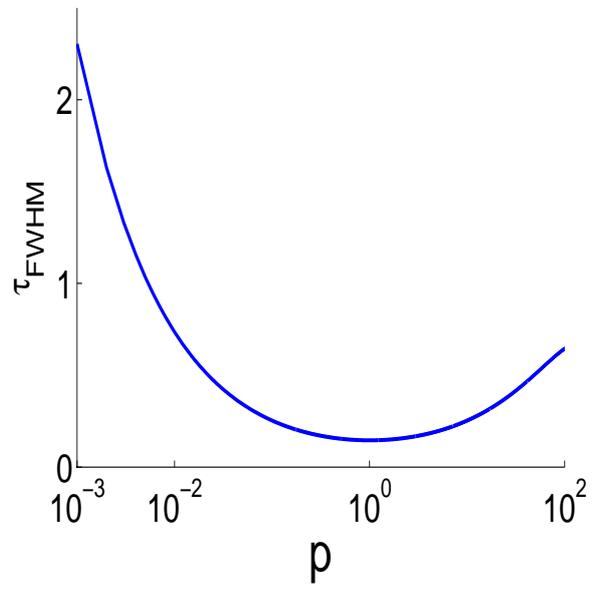}}
\caption{Variation of spatial width ($\tau_{FWHM}$) of solitons with
power $ p$.\ Figure shows existence of bistable solitons. }
\label{fig:fig3}
\end{figure}

\newpage
\begin{figure}
\centering \scalebox{0.5}
{\includegraphics[height=6.5in,width=6.5in]{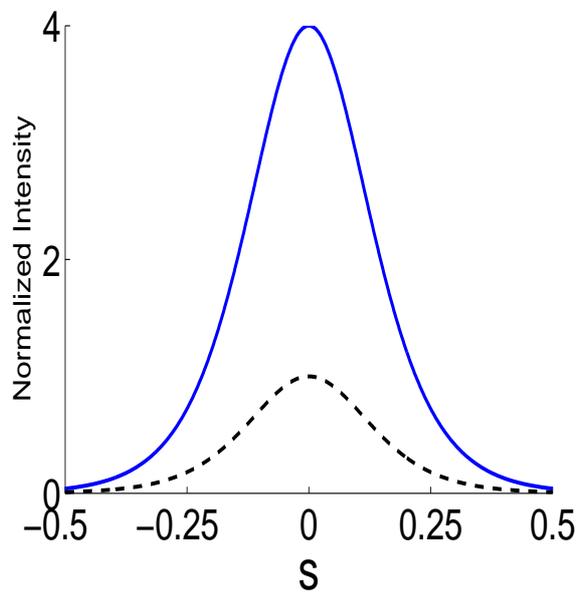}}
\caption{Normalized intensity  profile of a pair of bistable
solitons.    Both solitons have same spatial width but they possess
different peak power. } \label{fig:fig4}
\end{figure}

\begin{figure}
\centering \scalebox{0.5}
{\includegraphics[height=6.5in,width=6.5in]{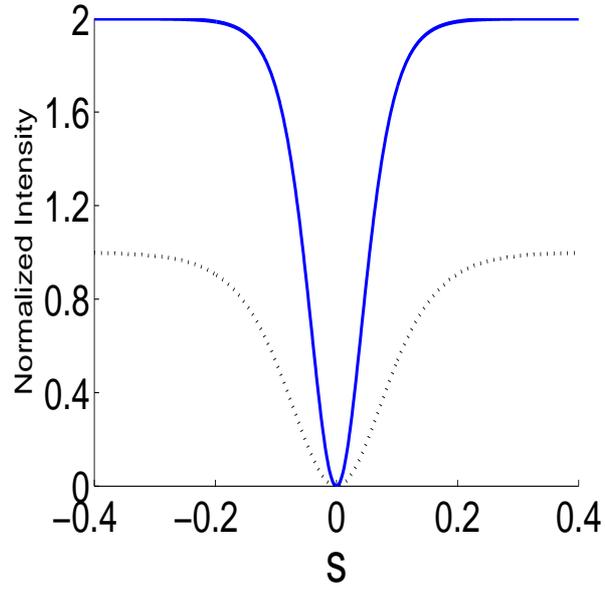}}
\caption{Normalised intensity profile of  dark spatial solitons for
$\beta=-43,  x_{0}=20\mu m; \rho=1$  and   $2$.} \label{fig:fig5}
\end{figure}

\begin{figure}
\centering \scalebox{0.5}
{\includegraphics[height=6.5in,width=6.5in]{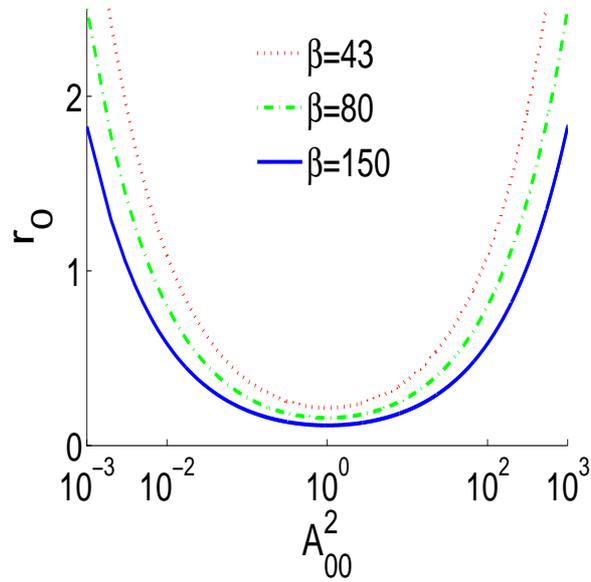}}
\caption{Existence curve of stationary solitons for different
$\beta$.} \label{fig:fig6}
\end{figure}

\begin{figure}
\centering \scalebox{0.5}
{\includegraphics[height=6.5in,width=6.5in]{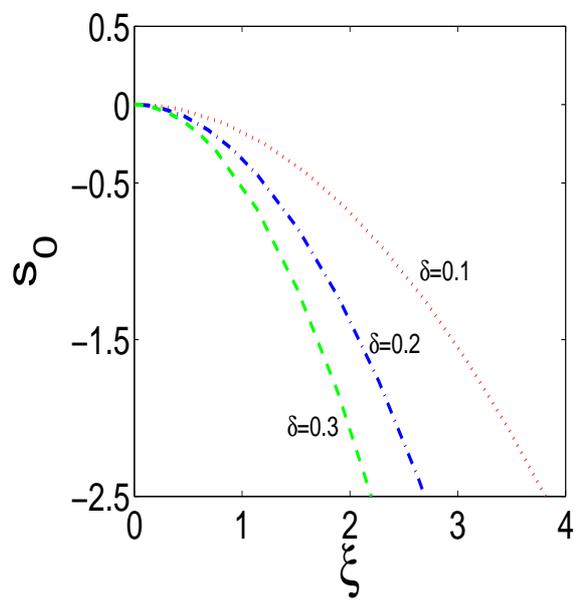}}
\caption{Spatial shift of soliton centre as it propagates through
the photovoltaic crystal.    $A_{00}=0.42$ and $r_{0}=0.298$. }
\label{fig:fig7}
\end{figure}

\begin{figure}
\centering \scalebox{0.5}
{\includegraphics[height=16.5in,width=16.5in]{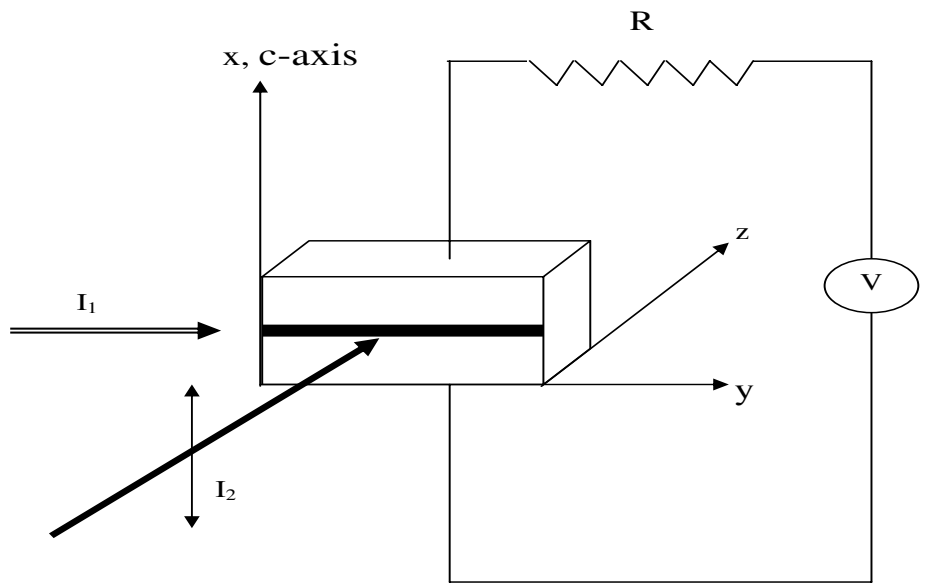}}
\caption{Optical configuration for two-photon photorefractive
effect. The crystal is illuminated with a gating beam of constant
intensity  $I_{1}$. } \label{fig:fig8}
\end{figure}

\begin{figure}
\centering \scalebox{0.5}
{\includegraphics[height=6.5in,width=6.5in]{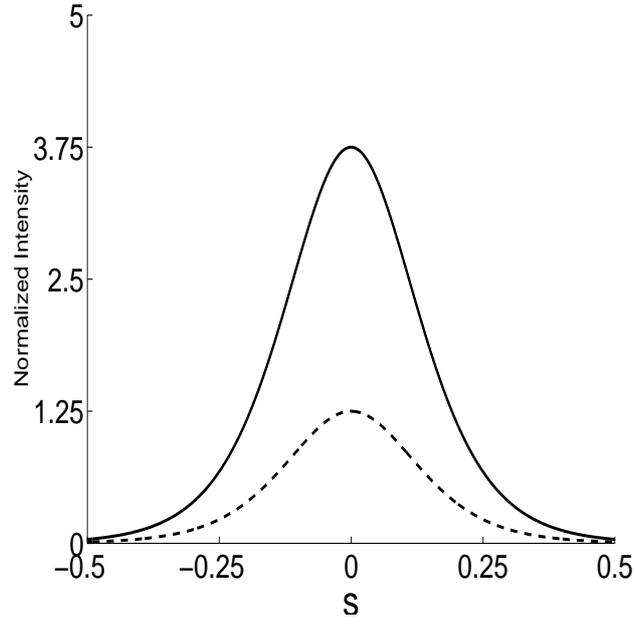}}
\caption{Components of a bright-bright soliton pair.    $\beta=43,
p=5 $    and   $\theta=30^{0}$. } \label{fig:fig9}
\end{figure}

\begin{figure}
\centering \scalebox{0.5}
{\includegraphics[height=6.5in,width=6.5in]{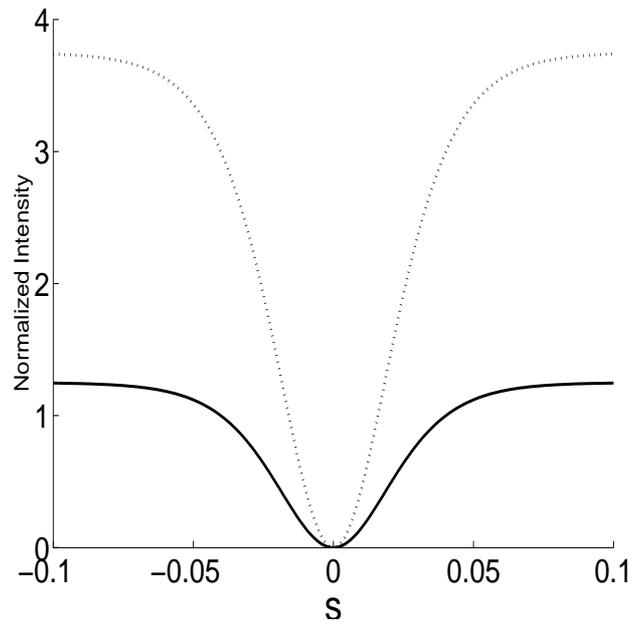}}
\caption{Components of a dark-dark soliton pair.        $\beta=-43,
\rho=5 $    and $\theta=30^{0}$. } \label{fig:fig10}
\end{figure}

\begin{figure}
\centering \scalebox{0.5}
{\includegraphics[height=6.5in,width=6.5in]{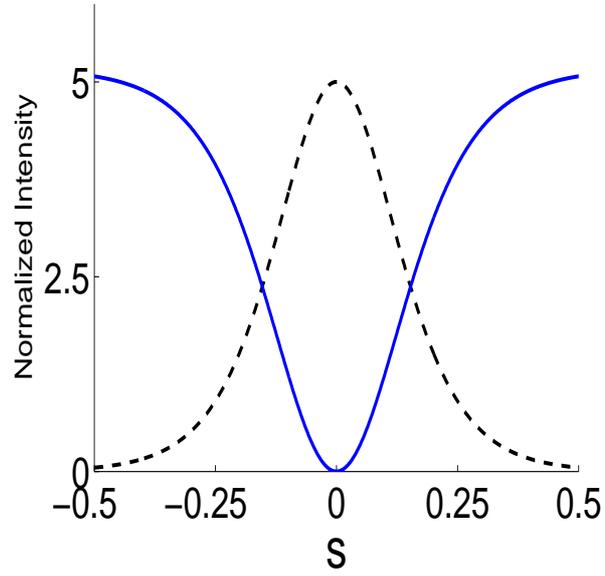}}
\caption{Components of a bright-dark soliton pair.     $\beta =-43,
\rho=5,   \Lambda=-0.01 $    and   $\theta=30^{0}$. }
\label{fig:fig11}
\end{figure}

\begin{figure}
\centering \scalebox{0.5}
{\includegraphics[height=6.5in,width=6.5in]{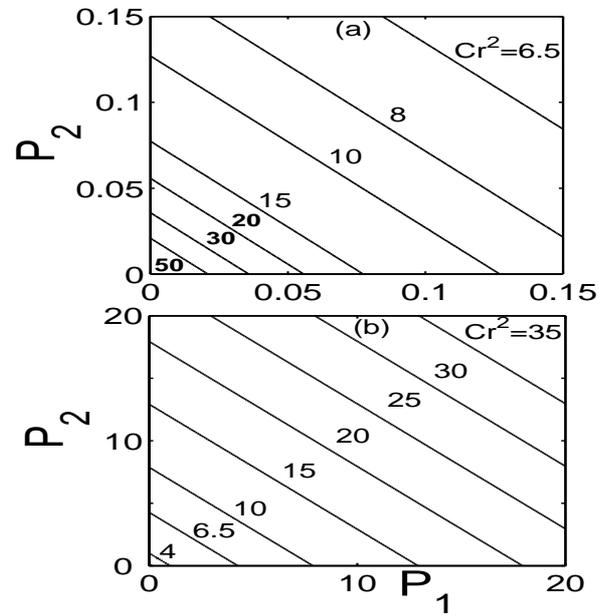}}
\caption{Existence curve  of screening photovoltaic solitons.  (a)
Low power, (b) High power. } \label{fig:fig12}
\end{figure}

\begin{figure}
\centering \scalebox{0.5}
{\includegraphics[height=6.5in,width=6.5in]{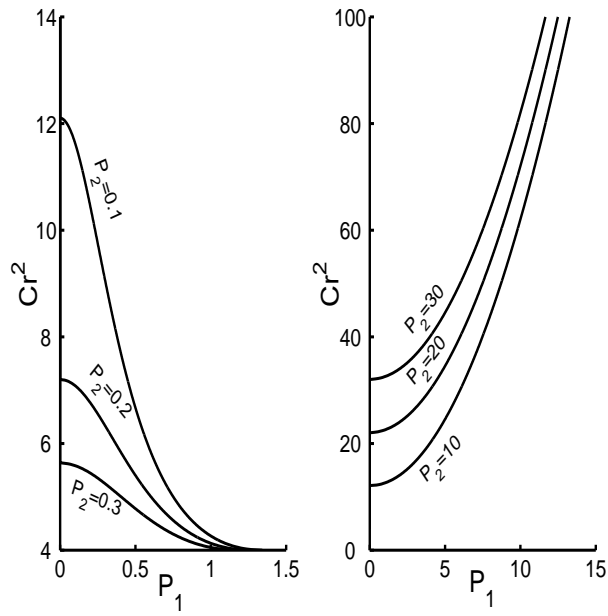}}
\caption{Variation of   peak power  $P_{1}$    of  one of the
component of the composite soliton    with spatial width $r$   while
the peak power of other component $P_{2}$ is constant. }
\label{fig:fig13}
\end{figure}

\begin{figure}
\centering \scalebox{0.5}
{\includegraphics[height=6.5in,width=6.5in]{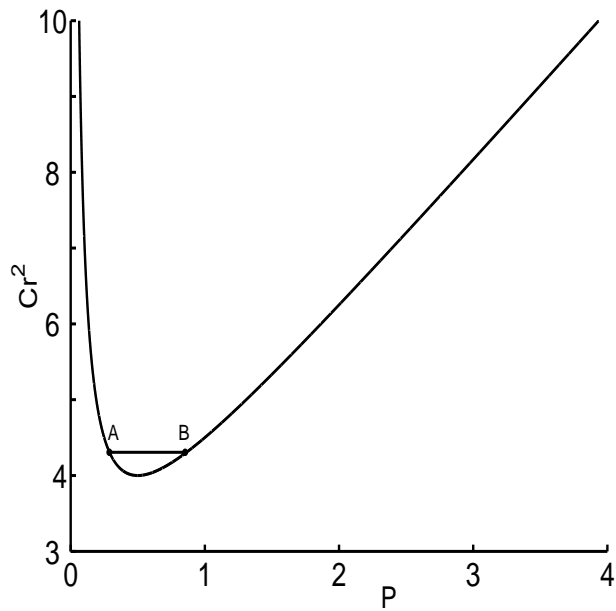}}
\caption{Variation of   peak power  $P$   of the degenerate
composite soliton   with spatial width $r$.  Nature of the curve
signifies existence of bistable property of solitons. }
\label{fig:fig14}
\end{figure}

\begin{figure}
\centering \scalebox{0.5}
{\includegraphics[height=6.5in,width=6.5in]{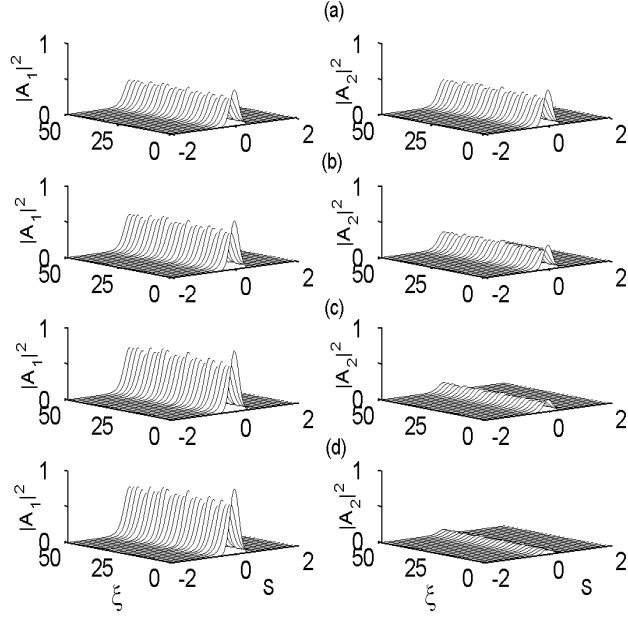}}
\caption{Stable propagation of composite solitons as obtained by
direct numerical    simulation.  $Cr^{2}=4$,  $r= 1/5$.   (a)
$P_{1}=0.5,P_{2}=0.5,$ (b) $ P_{1}=0.6666,P_{2}=0.3333,$   (c)
$P_{1}=0.8333,P_{2}=0.1666$, and (d)  $P_{1}=0.909,P_{2}=0.909$ .
Left panel $|A_{1}|^{2}$ and  right panel   $|A_{2}|^{2}$.}
\label{fig:fig15}
\end{figure}

\begin{figure}
\centering \scalebox{0.5}
{\includegraphics[height=6.5in,width=6.5in]{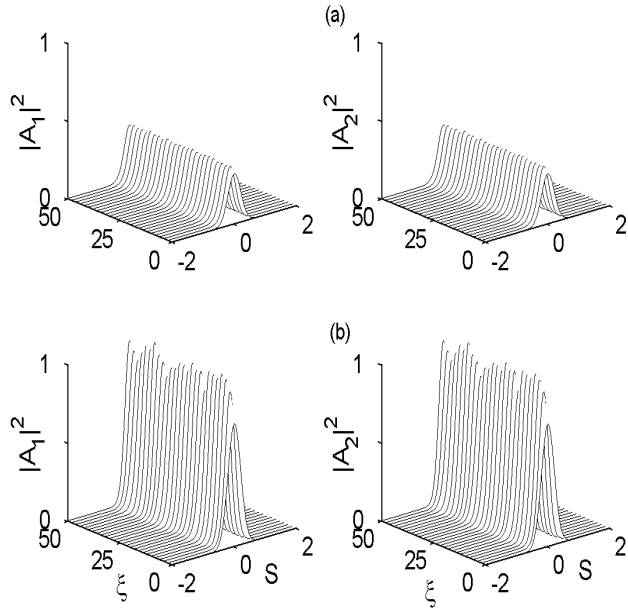}}
\caption{Propagation of bistable composite solitons. Peak power and
width of these solitons have been chosen from point A and B of
figure (14) which corresponds to $Cr^{2}=4$. Upper panel corresponds
to point A and lower panel corresponds to point B.  Solitons of both
upper and lower panels have same width i.e., each component has
spatial width $r=0.290$.  Peak power of each component in the upper
panel  $P=0.321$.  Peak power of each component in the  lower panel
$P=0.779$. } \label{fig:fig16}
\end{figure}

\begin{figure}
\centering \scalebox{0.5}
{\includegraphics[height=6.5in,width=6.5in]{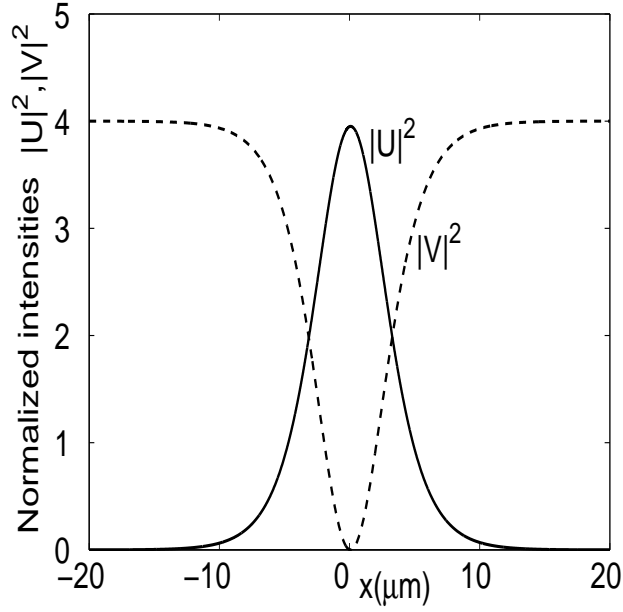}}
\caption{Soliton components $|U|^{2}$ and $|V|^{2}$ of the
bright-dark soliton pair for $\rho=4,   p=3.95$,     $\sigma=10^{6}$
and $\alpha=-22.2$. Gating beam intensity  $I_{1}=10^{6} W/m^{2}$
and calculated value of $\Delta=-0.01$. } \label{fig:fig17}
\end{figure}

\begin{figure}
\centering \scalebox{0.5}
{\includegraphics[height=6.5in,width=6.5in]{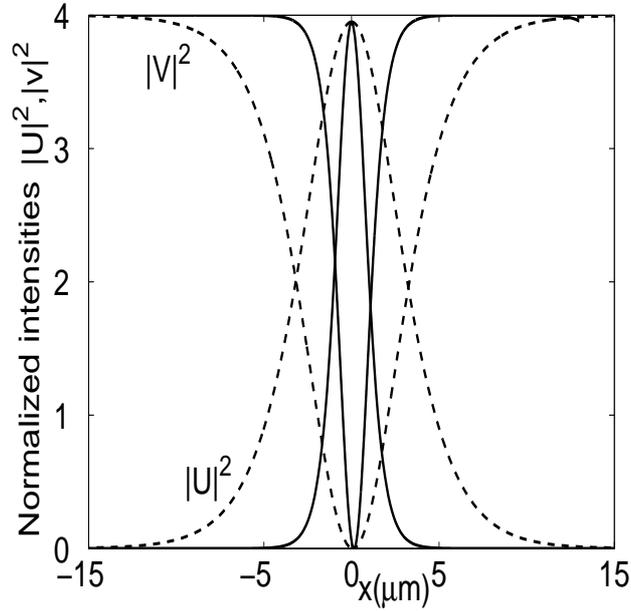}}
\caption{Soliton components $|U|^{2}$  and $|V|^{2}$  of the
bright-dark soliton pair for two different values of the gating beam
intensity $I_{1}$. Values of different parameters are
$\rho=4,p=3.95, \sigma=10^{6}$  and  $\alpha=-22.2$. Solid line for
$ I_{1}=10^{5} W/m^{2}$, dashed line for $I_{1}=10^{6} W/m^{2}$. }
\label{fig:fig18}
\end{figure}

\FloatBarrier
\begin{figure}
\centering \scalebox{0.5}
{\includegraphics[height=8.5in,width=8.5in]{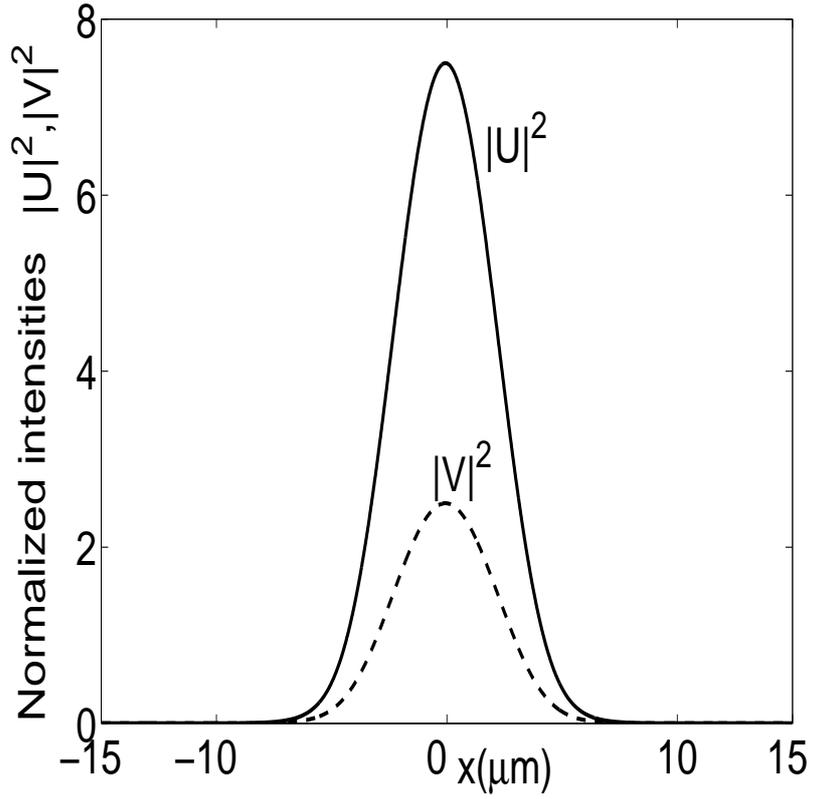}}
\caption{Soliton components   $|U|^{2}$   and $|V|^{2}$  of the
bright-bright soliton pair when  $p=10$  and $\theta=30^{0}$. }
\label{fig:fig19}
\end{figure}

\newpage
\begin{figure}
\centering \scalebox{0.5}
{\includegraphics[height=8.5in,width=8.5in]{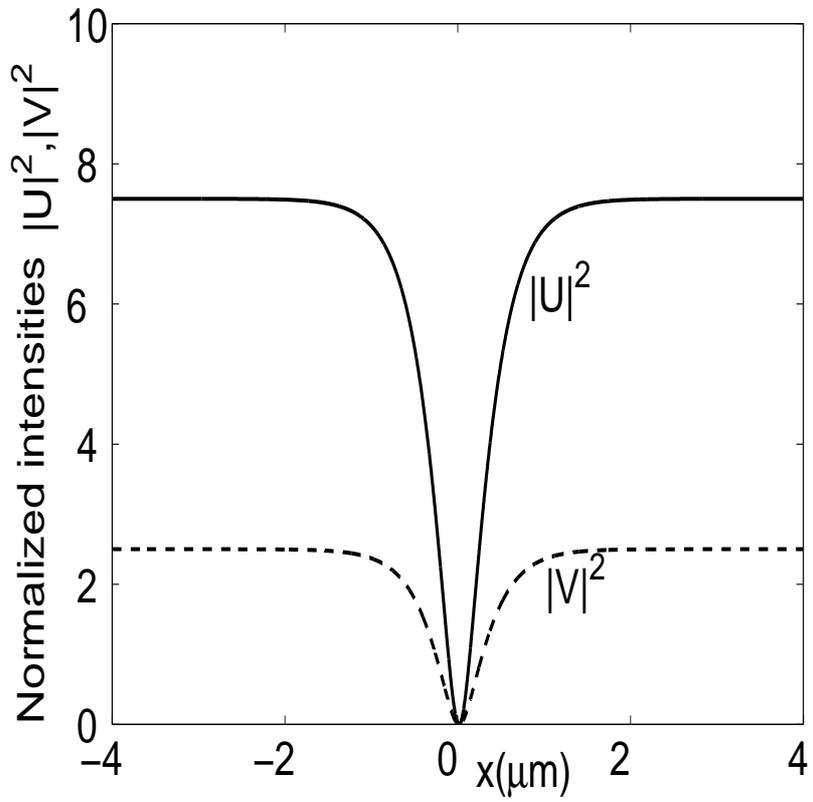}}
\caption{Soliton components $|U|^{2}$ and $|V|^{2}$  of the
dark-dark soliton pair when  and $\rho=10$  and $\theta=30^{0}$. }
\label{fig:fig20}
\end{figure}

\newpage
\addcontentsline{toc}{chapter}{List of Figures}
\begin{doublespacing}
\listoffigures
\end{doublespacing}


\newpage

\begin{thebibliography}{palinnat}

\bibitem[ Hasegawa et al.(1973)]{1}A. Hasegawa and F. Tappert,   Appl. Phys. Letts.  23(1973)142-145.\
%
\bibitem[Mollenaue et al.(1980)]{2}L. F. Mollenaue,  R. H. Stolen and J. P. Gordon, Phys. Rev.
Letts. 45(1980)1095.\
%
\bibitem[Agrawal et al.(1989)]{3}G. P. Agrawal,   Academic Press, New York
(1989).\
%
\bibitem[ Zabusky et al.(1965)]{4}N.J.Zabusky and M.D. Kruskal, Phys. Rev. Letts. 15(1965)240.\
%
\bibitem[Kivshar et al.(2001)]{5}Y.S. Kivshar  and A. A. Sukhorukov, in Spatial Solitons, S. Trillo and
W. Toruellas, Eds.( Springer, New York, 2001) pp 211-246.\
%
\bibitem[Akhmediev et al.(1997)]{6}N. N. Akhmediev and A. Ankiewicz, Solitons: Nonlinear Pulses and Beams (Chapman and Hall, London  1997, Chapter 3).\
%
\bibitem[Malomed et al.(2005)]{7} B. A. Malomed, D. Mihalache, F. Wise and L. Torner;   J. Opt. B:
Quantum Semiclass. Opt. 7(2005)R53-R72.\
%
\bibitem[ Kivshar et al.(2003)]{8}Y. S. Kivshar  and G. P. Agrawal, Optical Solitons: From Fibers to
Photonic Crystals,  Academic Press, San Diogo, California, 2003.\
%
\bibitem[Mihalache et al.(2008)]{9}D. Mihalache, D. Mazilu, F. Lederer, H. Leblond  and B. A. Malomed;
Physical Review A  77(2008)033817.\
%
\bibitem[Crosignani et al.(1993)]{10} B. Crosignani, M. Segev, D. Ergin, P. Di Porto,  A. Yariv  and G. Salamo, J. Opt. Soc. Am B 10(
1993)446-453.\
%
\bibitem[L$\ddot{u}$ et al.(2008) ]{11}  X. L$\ddot{u}$, Hong-Wu Zhu, Zhen-Zhi Yao, Xiang-Hua Meng, Cheng Zhang, Chun-Yi Zhang, B. Tian;   Annals of Physics  323, (2008)1947-1955.\

\bibitem[Tian et al.(2011)]{12}B.Tian and Jixiong Pu;  Optics Letters 36(2011)2014-2016\

\bibitem[L$\ddot{u}$ et al.(2008)]{13}X. L$\ddot{u}$, B. Tian, Tao Xu, Ke-Jie Cai, Wen-Jun Liu;   Annals of Physics 323(2008)2554-2565. \

\bibitem[Liu et al.(2010)]{14} W. J. Liu, B. Tian, Tao Xu, Kun Sun, Yan Jiang; Annals of Physics 325(2010)1633-1643. \

\bibitem[Tienman et al.(2008)]{15}M. Tiemann,  J. Petter, T. Tschudi, Optics Commun. 281(2008)175-180.\

\bibitem[Tienman et al.(2009)]{16}M. Tiemann, T. Halfmann, T. Tschudi, Opt. Commun. 282(2009)3612-3619.\

\bibitem[Anjan Biswas et al.(2006 )]{17} Anjan Biswas and Swapan Konar; Non-Kerr Law Optical
Solitons, CRC Press, New York (2006).\

\bibitem[Duree et al.(1993)]{18}G. C. Duree, J. L. Sultz, G. J. Salamo, M. Segev, A. Yariv,
B.Crosignani, P. Di Porto,  E. J. Sharp, R. R. Neurgaonkar,
Phys.Rev. Letts.  71(1993)533-536.\

\bibitem[Karamzin et al.(1975)]{19}Y. N. Karamzin,   A. P. Sukhorukov,  Soviet Physics JETP 41(1975)414.\

\bibitem[Sukhorukov et al.(2000)]{20}A. A. Sukhorukov, Phys. Rev.  E 61(2000)4530.\

\bibitem[Mihalache et al.(1997)]{21}D. Mihalache,  D. Mazilu, L. C. Crasovan and L.Toner,   Optics Commun.   137(1997)113.\

\bibitem[Darmanyan et al.(1998)]{22} S. Darmanyan, A. Kobyakov and F. Lederer,  Phys. Rev. E 57(1998)2344.\

\bibitem[Sirtori et al.(1992)]{23} C. Sirtori, F. Capsso, D.L. Sivco and A.Y. Cho, Phys. Rev. Letts.
68(1992)1010.\
\bibitem[Schmidt  et al.(1996)]{24}H. Schmidt and A. Imamoglu, Opt. Letts.   21(1996)1936.\
%
\bibitem[Yang  et al.(2009)]{25} X. Yang, S. Li, C. Zhang and H. Wang,  J. Opt Soc. Am. B 26(2009)1423.\
%
\bibitem[Doyeol et al.(1987)]{26}A. Doyeol and S.L. Chuan, IEEE J. Quntum Electron. QE-23(1987)2196.\
%
\bibitem[Rosencher et al.(1991)]{27}E. Rosencher and P. Bois, Phys. Rev. B 44(1991)11315.\
%
\bibitem[Chauvet et al.(2001)]{28}M.Chauvet, S.Chauvin, H.Maillotte,    Opt.  Letts.  26(2001)1344.\
%
\bibitem[Bosshard et al.(1994)]{29}C. Bosshard, P. V. Mamyshev, G. I. Stageman,  Opt. Letts. 19(1994)90.\
%
\bibitem[Chauvet et al.(1997 )]{30}M.Chauvet, S.A.Hawkins, G. Salamo, M. Segev, F. D. Bliss, G. Bryant,  Appl.  Phys. Letts. 70(1997)2499.\
%
\bibitem[Fazio et al.(2004)]{31}E. Fazio, F. Renzi, R. Rinaldi, M. Bertolotti, M. Chauvet, W. Ramadan, A. Petris, and V. I. Vlad, \ \ \ \ \ Appl. Phys.
Letts.    85(2004)2193-2195.\


\bibitem[Kuroda et al.(2002)]{32}K. Kuroda,  in  Progress in Photorefractive Nonlinear Optics, Eds Kazuo Kuroda, Taylor and
Francis New York (2002).\

\bibitem[Chen  et al.(1968 )]{33}F. S. Chen, J. T. LaMacchia and D. B.Fraser,   Appl. Phys. Letts.  13(1968)223.\

\bibitem[Yeh et al.(1993 )]{34}P. Yeh; Introduction to photorefractive nonlinear optics,  Wiley 1993.\

\bibitem[Gunter  et al.(2006 )]{35}P. Gunter and J.P.Huignard (Eds), Photorefractive Materials and Their Applications, Springer (2006).\

\bibitem[Ashkin  et al.(1996)]{36}A. A.Ashkin, G. D. Boyd, J. M. Dziedzic, R.G. Smith, A. A. Ballmann,  H.
J. Levinstin and K. Nassau,   Appl. Phys. Letts. 9(1966)72.\
%
\bibitem[Meerhotz  et al.(1994)]{37}K. Meerhotz, B. L. Volodin, S. B. Kippelen and N. Peyghambarian,  Nature 371(1994)497 .\
%
\bibitem[Moerner et al.(1994)]{38}W. E. Moerner and S. M. Silence,   Chem. Rev. 94(1994)127.\
%
\bibitem[Shih et al.(1999)]{39} M.F.Shih and F.W. Sheu,  Optics Letts.  24(1999)1853.\
%
\bibitem[Sheu et al.(2001)]{40}F. W.Sheu and M. F. Shih, J. Opt. Soc. Am. B   18(2001)785.\
%
\bibitem[Kukhtarev et al.(1979)]{41}N. V. Kukhtarev, V.B. Markov, S. G. Odulov, M. S. Soskin and V.L. Vinetskii,   Ferroelectrics   22(1979)  949.\
%
\bibitem[DelRe et al.(2001)]{42}E. DelRe, A. Ciattoni and  A. J. Agranat,Opt. Letts.  260(2001)908-910.\
%
\bibitem[Saffman et al.(1998)]{43}M. Saffman and A. A. Zozulya, Opt.Letts.   23(1998)1579.\
%
\bibitem[Liu et al.(1999)]{44}J. S. Liu and K. Q. Lu,   J. Opt. Soc. Am B  16(1999)550-555.\
%
\bibitem[Carvalho et al.(1996)]{45}M. I. Carvalho, S. R. Singh, D. N. Christodoulides, Opt.Commun. 124(1996)642.\
%
\bibitem[Christodoulides et al.(1995)]{46}D.  N. Christodoulides and M. I. Carvalho, J. Opt. Soc.Am B  12(1995)1628-1633.\
%
\bibitem[Segev et al.(1994)]{47} M. Segev, G. C. Valley, B. Crosignani, P. Di Porto, and A. Yariv,  Phys. Rev. Letts. 73(1994)3211.\
%
\bibitem[Shih et al.(1995)]{48}M.F.Shih, M. Segev, G.C. Valley, G.  Salamo, B.   Crosignani,P. Di Porto, Electron. Lett. 31(1995)826.\
%
\bibitem[Chen et al.(1996)]{49}Z. Chen,  M. Mitchell, M.F. Shih, M Segev,  M H.Garrett,   G. C. Valley, Opt. Letts.   21(1996)629.\
%
\bibitem[Iturbe-Castillo et al.(1994)]{50} M. D. Iturbe-Castillo, P. A. Aguilar,   J. J.S\'{a}nchez-Mondrag\'{o}n, S. Stepanov, V.  Vysloukh,
Appl. Phys. Letts. 64(1994)408\
%
\bibitem[Garca-Quirino et al.(1997)]{51} G. S. Garca-Quirino, M. D. Iturbe-Castillo, V. A. Vysloukh, J.J. S\'{a}nchez-Mondrag\'{o}n, S. I. Stepanov, G.Lugo-Martnez  and G. E. Torres-Cisneros, Opt.  Letts.   22(1997)154\
%
\bibitem[Mamaev et al.(1996)]{52} A.V. Mamaev, M. Saffman  and A. A. Zozulya,  Europhys.  Letts.
35(1996)25\
%
\bibitem[Mamaev et al.(1996)]{53}  A. V. Mamaev and M. Saffman,  Phys.  Rev. Letts. 76(1996)2262\
%
\bibitem[Mamaev et al.(1997)]{54}  A. V. Mamaev, A. A. Zozulya, V. K. Mezentsev, D. Z. Anderson
and M. Saffman,  Phys. Rev. A  56(1997)R1110-R1113.\
%
\bibitem[Krolikowski et al.(1998)]{55} W. Krolikowski, M. Saffman, B. Luther-Davies, and C.Denz, Phys.  Rev. Letts.  80(1998)3240.\
%
\bibitem[Konar et al.(2011)]{56} S.Konar, Phys. Express  1(2011)139.\
%
\bibitem[Valley et al.(1994)]{57}G. C. Valley, M. Segev, B. Crosignani, A. Yariv, M. M. Fejer and M. C. Bashaw, Phys. Rev. A 50(1994)R4457 .\
%
\bibitem[Taya et al.(1995)]{58}M. Taya, M. C. Bashaw, M. M. Fejer, M. Segev and G.C. Valley, Phys. Rev. A 52(1995)3095.\
%
\bibitem[Chen et al.(1997)]{59}Z. Chen, M. Segev, D. W. Wilson, R. E.Muller and P. D. Maker, Phys. Rev. Letts. 78(1997)2948.\
%
\bibitem[Liu et al.(1999)]{60}J.S.Liu and K.Q. Lu, J.Opt. Soc. Am B 16,(1999)550.\
%
\bibitem[Segev et al.(1997)]{61}M. Segev, G. C. Valey, M.C. Bashaw, M. Taya and M. M. Fejer;   J. Opt. Soc. Am. B 14,(1997) 1772.\
%
\bibitem[She et al.(1999)]{62}W. L. She, K. K. Lee and W. K. Lee,  Phys. Rev. Letts.   83(1999)3182 .\
%
\bibitem[Couton et al.(2004)]{63}G. Couton, H.Maillotte, M. Chauvet, J. Opt. B: Quantum Semiclass. Opt. 6(2004)S223.\
%
\bibitem[Konar et al.(2007)]{64}S. Konar, S. Jana,   S. Shwetanshumala, Opt. Commun. 273(2007)324.\
%
\bibitem[Vakhitiov et al.(1973)]{65}N. G.Vakhitiov and A. A. Kolokolov, Sov. Radio Phys. 16(1973) 783.\
%
\bibitem[DeAngelis et al.(1994)]{66}C. DeAngelis, IEEE J. QE 30,(1994) 818.\
%
\bibitem[Kumar et al.(1996)]{67}A.Kumar, T. Kurz and W. Lauterborn, Phys. Rev. E 53(1996)1166.\
%
\bibitem[Kaplan et al.(1985)]{68}A. E. Kaplan, Phys. Rev. Letts. 55(1985) 1291.\
%
\bibitem[Hasegawa et al.(1973)]{69}A. Hasegawa and F. Tappert, Appl. Phys. Letts. 23(1973)171.\
%
\bibitem[Krolikowski et al.(1996)]{70}W. Krolikowski, N. Akhmediev, B. Luther Davies, M. C. Golomb, Phys. Rev. E 54, (1996)5761-5765.\
%
\bibitem[Carvalho et al.(1995)]{71}M. I. Carvalho, S. R. Singh and D. N. Christodoulides, Opt. Commun. 120(1995)311.\
%
\bibitem[Carvalho et al.(2007)]{72}M. I.Carvalho, M. Facco  and D. N. Christodoulides, Phys Rev E 76(2007)016602.\
%
\bibitem[Zhan et al.(2010)]{73}K. Zhan, C.F. Hou, Yanwei Du, Opt. Commun. 283(2010)138-141.\
%
\bibitem[Zhang et al.(2008)]{74}G.Zhang, J.S.Liu, W. Cheng, Z.Huilan and  S.Liu,   Optik 119(2008) 3003-3008.\
%
\bibitem[Liu et al.(2007)]{75}S.Liu, J.S. Liu,  H. Zhang, G.Zhang, W. Cheng,    J.Modern Optics 54(2007)2795-2805.\
%
\bibitem[Petter et al.(1999)]{76}J. Petter,  C. Weilnau, C. Denz, A. Stepken, F. Kaiser, Optics Commun. 170(1999)291.\
%
\bibitem[Zhang et al.(2009)]{77}G. Zhang, Y. Han, L.Tao, A. Zheng, Q. Du,   Optics and Laser Technology  41 (2009)596-600.\
%
\bibitem[Liu et al.(2001)]{78}J.S. Liu,  D. Zhang, Z Hao, J. Modern Opt. 48(2001)1803-1810.\
%
\bibitem[Zhang et al.(2006)]{79}G.Zhang, J.S.Liu, S.Liu, H.Zhang  and W. Cheng,  J. Opt. A: Pure Appl. Opt.  8(2006)442-449.\
%
\bibitem[Liu et al.(2003)]{80}J.S.Liu and  Z. Hao,  Chinese Physics 12 (2003)1124.\
%
\bibitem[Liu et al.(2002)]{81}J.S. Liu and Z. Hao,  J. Opt. Soc. Am B 19(2002)513.\
%
\bibitem[Jiang et al.(2009)]{82} Q. C.Jiang,  Y. L.Su, X. M.Ji,  Chinese J. Quantum  Electron. 26(2009)619-623.\
%
\bibitem[Kodama et al.(1987)]{83}Y. Kodama and A. Hasegawa,  IEEE J. QE  223(1987)510.\
%
\bibitem[Blow et al.(1988)]{84}K. J. Blow, N. J. Doran and D. Wood, J.Opt. Soc. Am B 5(1988)1301.\
%
\bibitem[Sodha et al.(2008)]{85}M. S. Sodha,  S. K.  Agrawal and  A. Sharma, J. Plasma Physics 74(2008)65-77.\
%
\bibitem[Konar et al.(1994)]{86}S. Konar, A. Sengupta,   J.Opt. Soc. Am. B  11 (1994) 1644.\
%
\bibitem[Sturman et al.(1992)]{87}B. I. Sturman and V. M.Fridkin,  The photovoltaic and photorefractive effects in non-centrosymmetric materials ( Philadelphia, Gordon and Breach 1992).\
%
\bibitem[Zhang et al.(2007)]{88}G. Zhang, J.S. Liu, H. Zhang, C. Wang, S.Liu, Optik 118 (2007)440-444.\
%
\bibitem[Liu et al.(2002)]{89}J.S. Liu and  Z. Hao, Chinese Physics 11(2002)254-259.\
%
\bibitem[Ramadan et al.(2003)]{90}W. Ramadan, E.Fazio, A.Mascioletti, F.Inam, R.Rinaldi,A.Bosco, V.I.Vlad, A.Petris  and  M.Bertolotti,  J. Opt. A: Pure Appl.
Opt. 5(2003)S432.\\
%
\bibitem[Castro-Camus et al.(2003 )]{91}E. Castro-Camus and L.F.Magana, Opt.  Letts. 28(2003)1129.\

\bibitem[Hou et al.(2005 )]{92}C. F. Hou, Y. B. Pei, Z. X. Zhou and X. D.Sun, Phys. Rev. A 71 (2005)053817.\

\bibitem[Hou et al.(2007 )]{93}C. F.  Hou, Y. Zhang, Y. Y. Jiang and Y.B. Pei, Opt. Commun. 273(2007) 544-548.\

\bibitem[Zhang et al.(2009 )]{94}  G. Zhang,  J.S. Liu,   J. Opt. Soc. Am.B  26 (2009)113.\

\bibitem[Konar et al.(2010 )]{95}  S.Konar, S. Shekhar  and W.P.Hong, Optics and Laser Technology 42(2010) 1294-1300 .\

\bibitem[Zhang et al.(2007 )]{96}Y.Zhang, C.F.Hou  and  S.X. Dong,  Chinese Physics 16 (2007)159.\
\bibitem[Jiang et al.(2010 )]{97}Q. Jiang, Y. Su and X. Ji,  Optica Applicata  Vol XL. No 2. (2010)481.\

\bibitem[Jiang et al.(2011 )]{98}Q. Jiang, Y. Su and X. Ji,   Optics and Laser Technology  43 (2011)91-94.\
\bibitem[Ji et al.(2010 )]{99}X. Ji, Q. Jiang, J. Yao, J.S. Liu, Optics and Laser Technology   42(2010)322-327.\
\bibitem[Jiang et al.(2011 )]{100}Q.Jiang, Y.Su, X. Ji,  Optik 122(2011)490-493.\

\bibitem[Su et al.(2010 )]{101}Y. L. Su, Q. C. Jiang, X.M.Ji, Chinese J. of Quantum Electron.  27(2010)331-335.\

\bibitem[Zhang et al.(2010 )]{102}G. Zhang, Y. Cheng, Z. Luo,  L. Tao and Q. Du, Optics Commun. 283(2010)335-339.\

\bibitem[Vlad et al.(2006 )]{103}V. I. Vlad, A. Petris, A. Bosco, E. Fazio and M. Bertolotti, J. Opt. A: Pure Appl. Opt. 8 (2006) S477- S482.\
\bibitem[Segev et al.(1995 )]{104}M. Segev, G. C. Valley, S. R. Singh, M.I. Carvalho, D. N. Christodoulides, Opt. Letts. 20(1995)1764.\
\bibitem[Carvalho et al.(1996 )]{105}M. I. Carvalho,  S. R. Singh,  D. N. Christodoulides.  R. I. Joseph, Phys. Rev E  53 (1996)R53.\
\bibitem[Lu et al.(2002)]{106}K. Lu, S.Qian, W.Zhao, Y.Zhang, Z. Wu, Optics Commun. 209 (2002)437-444.\

\bibitem[Lu et al.(2004 )]{107}K.Q. Lu, W. Zhao, Y. Yang, C.Sun, X. Bin, Y.Zhang  and J. Xu,  J. Opt. A: Pure Appl. Opt. 6 (2004)658-665.\
\bibitem[Christodoulides et al.( 1996)]{108}D.N. Christodoulides, S. R. Singh, M. I. Carvalho  and M. Segev, Appl. Phys. Letts. 68 (1996)1763.\

\bibitem[Chen et al.(1997 )]{109}Z. Chen, M. Segev, T.H. Coskun, D.N. Christodoulides  and Y.S. Kivshar, J. Opt. Soc. Am. B 14 (1997)3066.\
\bibitem[Chen et al.( 1996)]{110}Z. Chen,  M. Segev, T. H. Coskun, D. N. Christodoulides, Opt. Lett. 21 (1996)1436-1438.\

\bibitem[Chen et al.(1996 )]{111}Z. Chen, M. Segev, T. H. Coskun, D. N. Christodoulides, Y.S. Kivshar  and  V. V. Afanasjev, Opt. Letts. 21(1996)1821.\

\bibitem[Hou et al.(2002)]{112}C.F. Hou, Y.Jiang, B. Yuan, X. Sun, C. Du, S. Li, Optical Materials 19(2002)377-381.\


\bibitem[Lu et al.(2005)]{113}K.Q. Lu, W.Zhao, Y. Yang, G.Chen, J. Xu, Y. Zhang,  X. Hou, Optical Materials 27 (2005) 1845-1850.\
%
\bibitem[Chen et al.(1996)]{114}Z. Chen, M. Segev, T. H. Coskun and D. N. Christodoulides, Opt. Letts.  21(1996)1436.\
%
\bibitem[Zakery et al.(2004)]{115}A. Zakery and K.Keshavarz, J. Phys. D: Appl. Phys. 37(2004)3409-3418.\
%
\bibitem[Zakery et al.(2004)]{116}A. Zakery, A. Keshavarz,  Optik 115 (2004) 507-511.\
%
\bibitem[Hou et al.(2002)]{117}C.F. Hou, D.C.Guang, Abdurusul, S.Q.Li,  Chinese Phys. Letts. 19 (2002)63.\
%
\bibitem[Hou et al.(2001)]{118}C.F. Hou, L. Bin, X.D.Sun , Y.Y. Jiang  and X.K.Bin, Chinese Phys. 10 (2001) 310-314.\
%
\bibitem[Hou et al.(2001)]{119} C. Hou, Z.Zhou, B.Yuan and X.Sun,  Appl. Phys B 72 (2001)191-194.\
%
\bibitem[Lu et al.(2001)]{120}K.Q. Lu,Y.Zhan, T.Tang and B. Li,  Phys Rev E 64(2001)056603.\
%
\bibitem[Weilnau et al.(2001)]{121}C. Weilnau, W. Krolikowski, E. A. Ostrovskaya, M. Ahles, M. Geisser, G. McCarthy, C. Denz, Y. S. Kivshar and B. Luther
Davies, Appl. Phys B 72 (2001)723-727.\

\bibitem[Hong et al.(2008)]{122}W. P. Hong, J. Korean Physical Society    53 (2008) 3207-3212.\

\bibitem[Hou et al.(2004)]{123}C. Hou, Z. Zhou, X. Sun, Opt. Mater. 27 (2004)63.\

\bibitem[Gardner et al.(1967)]{124}C.S. Gardner, J.M. Greene, M. D. Kruskal, and R. M. Miura, Phys. Rev. Letts. 19 (1967)1095.\

\bibitem[Ablowitz et al.(1981)]{125}M. J. Ablowitz, H. Segur, Solitons and the Inverse Scattering Technique, SIAM, Philadelphia, 1981.\

\bibitem[Akhmanov et al.(1968)]{126}S.A. Akhmanov, A.P. Sukhorukov  and  R.V. Khokhlov,  Sov. Phys. USP.  10 (1968)609.\

\bibitem[Akhmanov et al.(1972)]{127}S.A. Akhmanov, A.P. Sukhorukov, R.V. Khokhlov, A.T. Arechi,  E.D. Shulz Dubois (Eds.), Laser Handbook, Vol. II, North Holland, Amsterdam, 1972, p. 1151.\
%
\bibitem[Vlasov et al.(1971)]{128}S.N.Vlasov, V.A.Petrischev  and V.I.Talanov, Sov. Radio Phys. 14(1971)1062.\
%
\bibitem[Anderson et al.(1983)]{129}D. Anderson,  Phys. Rev. A 27 (1983)3135.\
%
\bibitem[Malomed et al.(1997)]{130}B.Malomed, D. Anderson, M. Lisak and M. L. Quiroga-Teixeiro and L. Stenflo, Phys. Rev. E  55 (1997)962.\
%
\bibitem[Mihalache et al.(2010)]{131}D. Mihalache, D. Mazilu, V. Skarka, B. A. Malomed,3H. Leblond, N. B. Aleks$\acute{i}$c, and F. Lederer ; Physical Review A 82 (2010) 023813 \
%
\bibitem[Mihalache et al.(2007)]{132}D. Mihalache, D. Mazilu, F. Lederer, H. Leblond, and B. A. Malomed; Physical Review A  75, (2007) 033811 .\
%
\bibitem[Mihalache et al.(2010)]{133}D. Mihalache, D. Mazilu, V. Skarka, B.A.Malomed, H. Leblond, N.B.Aleks$\acute{i}$c, F. Lederer;  Phys. Rev. A 82 (2010)023813.\
%
\bibitem[Jana  et al.(2007)]{134}S. Jana and S. Konar,   Phys. Letts. A 362,  (2007)435-438.\

\bibitem[Anderson et al.(2001)]{135}D. Anderson, M. Lisak and A. Berntson, Pramana-journal of Physics  57 (2001)917.\

\bibitem[Zhang et al.(2008)]{136}Y. Zhang, C.F. Hou,  F.Wang, X.Sun,   Optik 119 (2008)700\

\bibitem[Lu et al.(2007)]{137}K. Lu, W.Zhao, Y. Yang, Y. Yang, M. Zhang, R. A. Rupp, M. Fally, Y. Zhang  and  J. Xu,  Appl. Phys. B 87(2007)469-473.\
%
\bibitem[Zhan et al.(2010)]{138}K. Zhan, C.F. Hou, H. Tian, S. Pu and Y Du,  J Opt.  12 (2010)015203.\
%
\bibitem[Zhan et al.(2010)]{139}K. Zhan,C.F. Hou and Y.Zhang,  J. Opt.  12 (2010)035208.\
%
\bibitem[Zhan et al.(2010)]{140}K.Zhan, C.F. Hou, T.Hao, Y.Zhang, Phys. Letts. A 374 (2010)1242.\

\bibitem[Jiang et al.(2010)]{141}Q. Jiang, Y. Su, X. Ji, Optics and laser Technology 42 (2010)720-723.\

\bibitem[Konar et al.(2010)]{142}S.  Konar and N. Asif, Phys. Scr. 81 (2010)015401.\

\bibitem[Asif et al.(2008)]{143}N. Asif, S. Shwetanshumala and S. Konar, Phys. Letts. A 372 (2008)735-740.\

\bibitem[Srivastava et al.(2009)]{144}S. Srivastava and S.Konar, Optics and Laser Technology 41 (2009)419-423.\

\bibitem[Su et al.(2010)]{145}Y. L. Su, Q. C. Jiang and X. M. Ji,  Commun. Theor. Phys. 53 (2010)943-946.\

\bibitem[Keshavarz et al.(2009)]{146}A. Keshavarz, L. Sadralsadati and M. Hatami, Progress In Electromagnetics Research Symposium Proceedings, Moscow, Russia, August 18-21, pp-1823(2009).\
%
\bibitem[Ji et al.(2012)]{147}Xuanmang Ji, Jinlai Wang, Qichang Jiang and Jinsong Liu;  Phys. Scr.
85 (2012) 025403.\
%
\bibitem[Zhang et al.(2011)]{148}Y.Zhang, C.F Hou, K. Zhan, X. Sun, Optik 122 (2011)263-265.\
%
\bibitem[Ji et al.(2010)]{149}S.H.Ji,  Laser Technology    34 (2010) 202 \
%
\bibitem[Zhan et al.(2010)]{150}K Y Zhan, C F Hou, H Tian, S Z Pu  and Y W Du;    J. Opt. 12(2010) 015203.\
%

\bibitem[Su et al.(2011)]{151}Y. Su, Q. Jiang, X. Ji and J. Wang, Optics and Lasers in Engineering 49(2011)526–529.\

\bibitem[Zhang et al.(2007)]{152}Y. Zhang, C. F. Hou, X. D. Sun, Acta.  Phys. Sin 56,(2007)3261.
%

%




\end{thebibliography}
\end{document}